\documentclass[reprint, superscriptaddress, twocolumn, amsmath, amssymb, aps, prb]{revtex4-2}
\usepackage{graphicx}
\usepackage{dcolumn}
\usepackage{bm}
\usepackage{placeins}
\usepackage{appendix}
\usepackage{upgreek}

\usepackage{tabularx}
\usepackage{array}

\usepackage{verbatim}
\usepackage[usenames,dvipsnames]{xcolor}
 \usepackage{amsmath}
 \usepackage{amsfonts}
 \usepackage{amssymb}

 \usepackage{bbold}     
 \usepackage[makeroom]{cancel}  
 \usepackage{multirow}    
 \usepackage[normalem]{ulem}        
 \usepackage{array}
 \usepackage{pdfpages}
\usepackage{romannum}
\makeatletter
 \AtBeginDocument{\let\LS@rot\@undefined}
 \makeatother

\begin{document}


\title{Strong enhancement of g-factor in PbTe-Pb hybrid nanowires}


\author{Shan Zhang }
\email{equal contribution}
\affiliation{State Key Laboratory of Low Dimensional Quantum Physics, Department of Physics, Tsinghua University, Beijing 100084, China}

\author{Wenyu Song}
\email{equal contribution}
\affiliation{State Key Laboratory of Low Dimensional Quantum Physics, Department of Physics, Tsinghua University, Beijing 100084, China}

\author{Zonglin Li }
\email{equal contribution}
\affiliation{State Key Laboratory of Low Dimensional Quantum Physics, Department of Physics, Tsinghua University, Beijing 100084, China}

\author{Zehao Yu}
\email{equal contribution}
\affiliation{State Key Laboratory of Low Dimensional Quantum Physics, Department of Physics, Tsinghua University, Beijing 100084, China}

\author{Ruidong Li}
\affiliation{State Key Laboratory of Low Dimensional Quantum Physics, Department of Physics, Tsinghua University, Beijing 100084, China}

\author{Yuhao Wang}
\affiliation{State Key Laboratory of Low Dimensional Quantum Physics, Department of Physics, Tsinghua University, Beijing 100084, China}

\author{Zeyu Yan}
\affiliation{State Key Laboratory of Low Dimensional Quantum Physics, Department of Physics, Tsinghua University, Beijing 100084, China}

\author{Jiaye Xu}
\affiliation{State Key Laboratory of Low Dimensional Quantum Physics, Department of Physics, Tsinghua University, Beijing 100084, China}

\author{Zhaoyu Wang}
\affiliation{State Key Laboratory of Low Dimensional Quantum Physics, Department of Physics, Tsinghua University, Beijing 100084, China}

\author{Yichun Gao}
\affiliation{State Key Laboratory of Low Dimensional Quantum Physics, Department of Physics, Tsinghua University, Beijing 100084, China}

\author{Shuai Yang}
\affiliation{State Key Laboratory of Low Dimensional Quantum Physics, Department of Physics, Tsinghua University, Beijing 100084, China}

\author{Lining Yang}
\affiliation{State Key Laboratory of Low Dimensional Quantum Physics, Department of Physics, Tsinghua University, Beijing 100084, China}

\author{Xiao Feng}
\affiliation{State Key Laboratory of Low Dimensional Quantum Physics, Department of Physics, Tsinghua University, Beijing 100084, China}
\affiliation{Beijing Academy of Quantum Information Sciences, Beijing 100193, China}
\affiliation{Frontier Science Center for Quantum Information, Beijing 100084, China}
\affiliation{Hefei National Laboratory, Hefei 230088, China}

\author{Tiantian Wang}
\affiliation{Beijing Academy of Quantum Information Sciences, Beijing 100193, China}
\affiliation{Hefei National Laboratory, Hefei 230088, China}

\author{Yunyi Zang}
\affiliation{Beijing Academy of Quantum Information Sciences, Beijing 100193, China}
\affiliation{Hefei National Laboratory, Hefei 230088, China}

\author{Lin Li}
\affiliation{Beijing Academy of Quantum Information Sciences, Beijing 100193, China}

\author{Runan Shang}
\affiliation{Beijing Academy of Quantum Information Sciences, Beijing 100193, China}
\affiliation{Hefei National Laboratory, Hefei 230088, China}

\author{Qi-Kun Xue}
\affiliation{State Key Laboratory of Low Dimensional Quantum Physics, Department of Physics, Tsinghua University, Beijing 100084, China}
\affiliation{Beijing Academy of Quantum Information Sciences, Beijing 100193, China}
\affiliation{Frontier Science Center for Quantum Information, Beijing 100084, China}
\affiliation{Hefei National Laboratory, Hefei 230088, China}
\affiliation{Southern University of Science and Technology, Shenzhen 518055, China}

\author{Ke He}
\email{kehe@tsinghua.edu.cn}
\affiliation{State Key Laboratory of Low Dimensional Quantum Physics, Department of Physics, Tsinghua University, Beijing 100084, China}
\affiliation{Beijing Academy of Quantum Information Sciences, Beijing 100193, China}
\affiliation{Frontier Science Center for Quantum Information, Beijing 100084, China}
\affiliation{Hefei National Laboratory, Hefei 230088, China}

\author{Hao Zhang}
\email{zhanghao@sustech.edu.cn}
\affiliation{State Key Laboratory of Low Dimensional Quantum Physics, Department of Physics, Tsinghua University, Beijing 100084, China}
\affiliation{Beijing Academy of Quantum Information Sciences, Beijing 100193, China}
\affiliation{Frontier Science Center for Quantum Information, Beijing 100084, China}
\affiliation{College of Semiconductors, Southern University of Science and Technology, Shenzhen 518055, China}

\begin{abstract}

We report large Land\'e g-factors observed in PbTe-Pb hybrid nanowires. The g-factor can reach 83, significantly larger than those in bare PbTe nanowires (typically below 20). We attribute this enhancement to orbital effects in the superconducting film, particularly when the magnetic field is nearly perpendicular to the Pb film. This enhancement is beneficial for the search for topological superconductivity by reducing the critical magnetic field required for the phase transition.

\end{abstract}

\maketitle  

Semiconductor nanowires coupled to superconductors have been extensively studied as a promising platform for the realization of Majorana zero modes \cite{Lutchyn2010, Oreg2010, Mourik, Deng2016, Gul2018, Song2022, WangZhaoyu, Delft_Kitaev, MS_2023, MS_2025, NextSteps, Prada2020, Leo_perspective}. A key parameter in this model is the electron Land\'e g-factor, which determines the critical magnetic field of the topological phase transition: $E_Z=\frac{1}{2}g\mu_BB>\sqrt{\mu^2+\Delta^2}$ \cite{Lutchyn2010, Oreg2010}. Here, $E_Z$ is the Zeeman energy, $\mu_B$ is the Bohr magneton, $B$ is the applied magnetic field, $\mu$ is the electrochemical potential, and $\Delta$ is the induced superconducting gap. Since external magnetic fields suppress superconductivity, a large g-factor is advantageous as it lowers the critical field of the topological phase transition. Previous studies have primarily focused on InAs and InSb nanowires, which possess large bulk g-factors reaching $\sim$40 \cite{Kammhuber2016}. However, when coupled to a superconductor (e.g., Al), this g-factor is significantly reduced due to the metallization effect \cite{Loss_metalization}. For example, in InAs-Al hybrid devices, the measured g-factor in tunneling spectroscopy experiments is typically around 3-6 \cite{Deng2016, WangZhaoyu, MS_2023}, resulting in a high critical field ($>$1 T) for the topological phase transition. To preserve the superconductivity of the parent superconductor, the Al film must be very thin, and the magnetic field orientation must be aligned with the film. A large g-factor allows the realization of Majoranas at low magnetic fields, thereby circumventing these constraints. Lower magnetic fields also alleviate the technical challenges in the design of field-compatible braiding circuits.

Here, we report large g-factors observed in PbTe-Pb hybrid nanowire devices. PbTe nanowires have recently been proposed and studied in depth as a potential Majorana candidate \cite{CaoZhanPbTe, Jiangyuying, Erik_PbTe_SAG, PbTe_AB, Fabrizio_PbTe, Zitong_JJ, Wenyu_QPC, Yichun_Gap, Yuhao_QPC, Ruidong_PJJ, Vlad_PbTe, Wenyu_Disorder, PbTe_In, Yuhao_degeneracy, Yichun_SQUID, Quantized_Andreev, Zonglin_Anisotropy}. For a nearly out-of-plane $B$, the g-factor in PbTe-Pb can reach 75-83, several times larger than that in PbTe nanowires. Based on the criterion above, this large g-factor can reduce the critical field of the topological phase transition to below 0.2 T for an induced gap of $\Delta$ = 0.4 meV. Notably, the parent superconductor Pb remains superconducting at this out-of-plane field \cite{Zonglin_Anisotropy}, paving the way for the possible realization of Majoranas in future studies.

\begin{figure}[b]
\includegraphics[width=0.95\columnwidth]{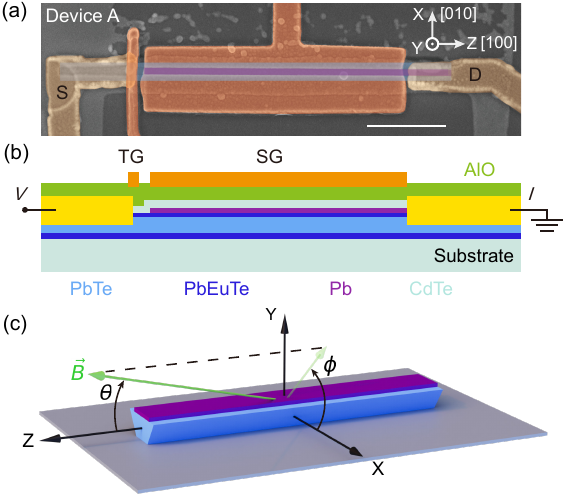}
\centering
\caption{Device set up. (a) SEM of device A. Scale bar is 1 micron. Source and drain contacts are denoted as S and D, respectively. (b) Schematic of the device. The Al$_2$O$_3$ dielectric is in green (labeled as AlO). (c) 3D illustration of the coordinate axes.  }
\label{fig2}
\end{figure}

Figure 1(a) shows a scanning electron micrograph (SEM) of a PbTe-Pb nanowire (device A). All transport data presented in Figures 2-4 were obtained from this device. In the Supporting Information, we also present data from a second device exhibiting similar characteristics. Details of the device growth and fabrication can be found in our previous works \cite{Wenyu_Disorder, Zonglin_Anisotropy}. The device schematic (not to scale) is illustrated in Fig. 1(b). Starting from a CdTe substrate, a Pb$_{0.99}$Eu$_{0.01}$Te buffer layer (dark blue, labeled as PbEuTe) was selectively grown. Subsequently, a 40-nm-thick PbTe nanowire (light blue) was grown selectively, followed by a 4-nm-thick Pb$_{0.97}$Eu$_{0.03}$Te interlayer (dark blue). This interlayer serves to tune the coupling between the nanowire and the Pb film (violet, 7-nm-thick) deposited on top. The device was then capped by a CdTe layer (cyan) \textit{in situ}. Finally, contacts (yellow) and top gates (orange) were fabricated. 

\begin{figure}[b]
\includegraphics[width=\columnwidth]{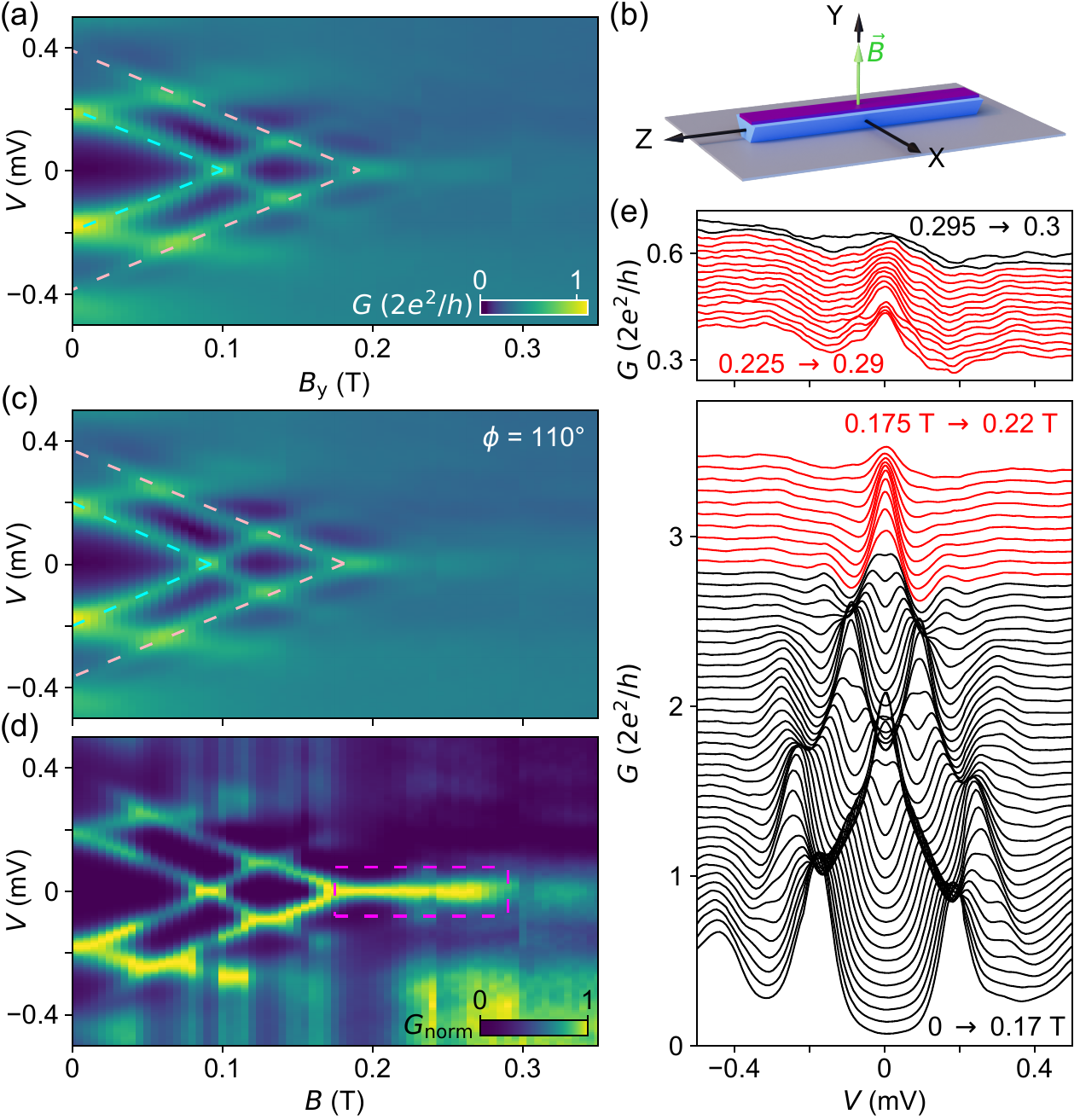}
\centering
\caption{(a) $G$ as a function of $V$ and $B$. $B$ is aligned with the $y$ axis as shown in (b). The cyan and pink dashed lines are linear fits of the subgap state dispersion. (c) $B$ scan by orienting $B$ at $\theta$ = 90$^{\circ}$, $\phi$ = 110$^{\circ}$, i.e. 20$^{\circ}$ away from the $y$ axis. The color bar is identical to that in (a). For (a) and (c), $V_{\text{TG}}$ = -3.57 V,  $V_{\text{SG}}$ = -3.496 V. (d) Replot of (c) by normalizing the conductance at each $B$. (e) Waterfall plot of (c), shown in two panels (for clarity). The vertical offsets are 0.07 (lower panel) and 0.02 (upper panel) in the unit of $2e^2/h$. The red curves are ZBPs.}
\label{fig2}
\end{figure}

The device geometry, a normal metal-nanowire-superconductor junction, is typical for tunneling spectroscopy experiments. The tunneling gate (TG) tunes the barrier region uncovered by Pb, while the SG tunes the PbTe region covered by Pb. Standard two-terminal measurements were performed in a dilution refrigerator at a base temperature below 50 mK. We measured the differential conductance, $G \equiv dI/dV$, as a function of $B$ and gate voltages ($V_{\text{TG}}$ and $V_{\text{SG}}$). The series resistance from the measurement setup such as fridge filters has been subtracted from both $G$ and the bias voltage $V$. No contact resistance was subtracted.

Since PbTe nanowires exhibit strong anisotropy \cite{Zonglin_Anisotropy}, the $B$ orientation is a crucial factor in studying the device transport. Figure 1(c) defines the coordinate system. The $z$ axis points along the nanowire, the $y$ axis is perpendicular to the substrate (out-of-plane), and the $x$ axis is in-plane and perpendicular to the nanowire. Furthermore, the angle between $B$ and the $z$ axis is denoted as $\theta$, while $\phi$ refers to the angle between the $x$ axis and the projection of $B$ onto the $xy$ plane.

The device was then tuned into the tunneling regime to study the evolution of subgap states as a function of $B$, as shown in Fig. 2(a). The gate settings are provided in the caption. The magnetic field was applied along the out-of-plane direction (Fig. 2(b)). For $B$ scans along the other two axes, see Figs. S1(e-f). Two subgap states located around $\pm$0.2 mV at zero field, exhibit Zeeman splitting with increasing $B$. The cyan dashed lines mark their lower-energy branch. The slope of these lines provides an estimation of the g-factor: $\Delta E = \frac{1}{2}g\mu_BB$, where $\Delta E$ corresponds to the energy shift in $V$. The extracted g-factor is approximately 68. A similar analysis can be applied to the outer subgap states (indicated by the orange dashed lines), yielding a g-factor of 70. In Fig. S2 of the Supporting Information, we show 6 additional $B$-scans at different gate settings, all revealing similarly large g-factors.

\begin{table}[b]
\caption{Summary of g-factors in PbTe nanowires. The magnetic field orientation is out-of-plane.}
\resizebox{\columnwidth}{!}{
\centering
\renewcommand\arraystretch{1.3}
\begin{tabular}{|c|c|c|c|c|}
\hline
Ref. & Substrate & Wire orientation & Device & g-factor \\
\hline\hline
\cite{Zonglin_Anisotropy} & CdTe(001) &  \{100\} & QD & 5-15 \\
\hline
\cite{Wenyu_Disorder} & CdTe(001) & \{100\} & QPC & 8-14 \\
\hline
\cite{Yuhao_degeneracy} & CdTe(001)  &\{100\}  &  QPC & 12-18 \\
\hline\hline
\cite{Yuhao_QPC} & CdTe(001) & \{110\} &  QPC & 23\\
\hline
\cite{Yuhao_QPC} & CdTe(110) & \{001\} &  QPC & 13-18\\
\hline
\cite{Wenyu_QPC} & CdTe(001) & \{110\} &  QPC & 8-10\\
\hline
\cite{Fabrizio_PbTe} & InP(111) & \{01$\overline{1}$\}, \{$\overline{1}2\overline{1}$\} &  QD & 1-22\\
\hline
\end{tabular}
}
\end{table}

\begin{figure*}[htb]
\includegraphics[width=\textwidth]{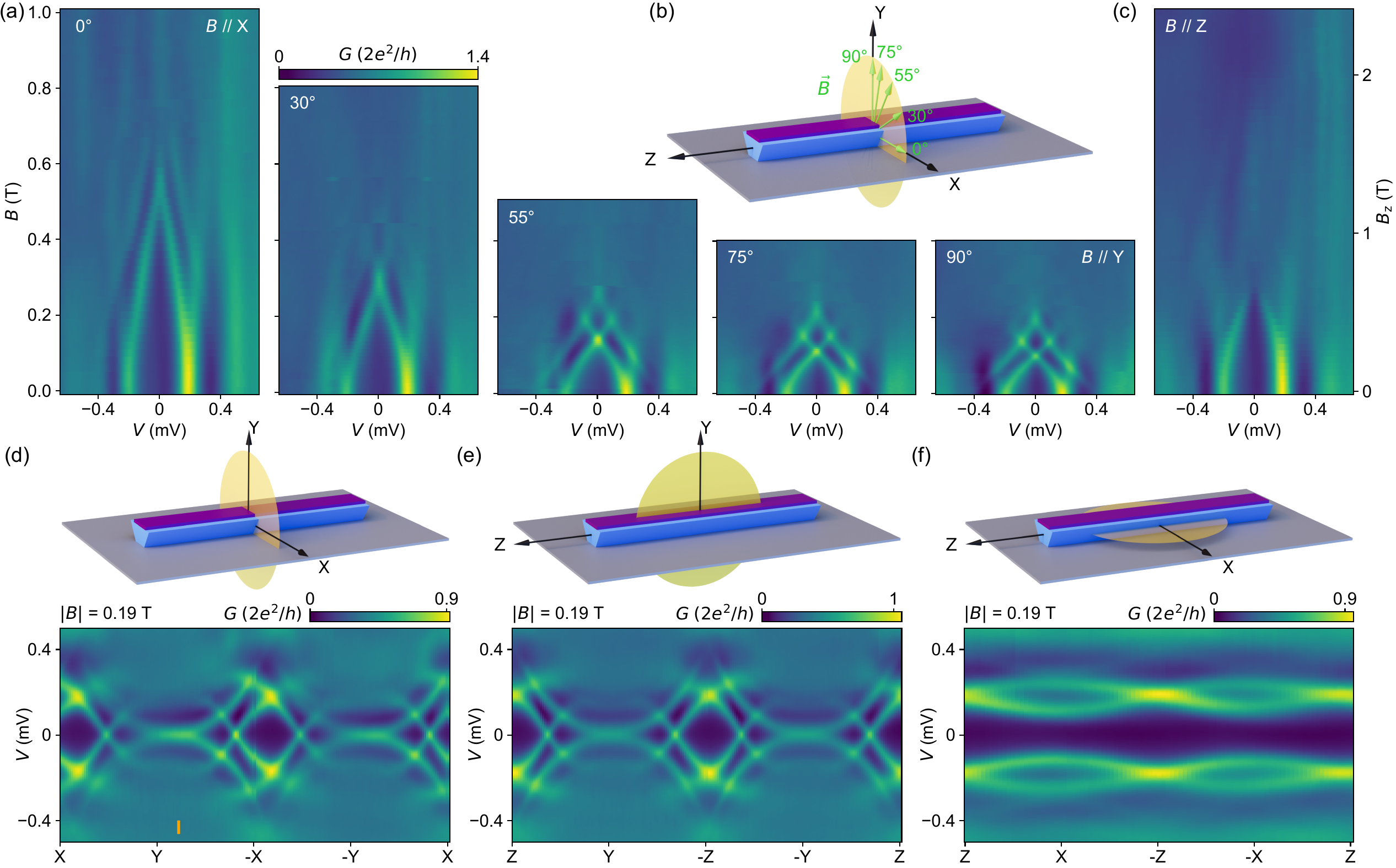}
\centering
\caption{(a) $B$ scans along different orientations in the $xy$ plane. $\theta$ = 90$^{\circ}$. From left to right, $\phi$ = 0$^\circ$, 30$^\circ$, 55$^\circ$, 75$^\circ$, 90$^\circ$, respectively. $V_{\text{TG}}$ = -3.714 V for the 0$^{\circ}$ and 90$^{\circ}$ cases, and -3.73 V for other panels.  $V_{\text{SG}}$ = -3.456 V for the 0$^{\circ}$ case, -3.444 V for the 90$^{\circ}$ case, and -3.48 V for the rest panels. (b) Schematic of these orientations, indicated by the arrows. (c) $B$ scan along the nanowire axis (the $z$ axis). The color bar is identical to that in (a). $V_{\text{TG}}$ = -3.714 V, $V_{\text{SG}}$ = -3.456 V. (d-f) Rotating $B$ in $xy$, $yz$, and $xz$ planes, respectively. $|B|$ is fixed at 0.19 T.  Upper panels are the schematics of the rotating planes. $V_{\text{TG}}$ = -3.57 V, $V_{\text{SG}}$ = -3.496 V. The minor adjustments in gate voltages are to compensate for the gate drift between scans.}
\label{fig2}
\end{figure*}

These values are significantly larger than those reported for bare PbTe nanowires. Table \Romannum{1} lists the extracted g-factors for PbTe nanowires from the literature. Our comparison focuses on the case where $B$ is out-of-plane. Given the strong anisotropy of PbTe, the crystal orientation and substrate also play critical roles. We thus restrict  our comparison to the first three rows (Ref. \cite{Zonglin_Anisotropy, Wenyu_Disorder, Yuhao_degeneracy}), where the substrate (CdTe(100)) and wire orientation (\{100\}) are identical to those of the PbTe-Pb devices reported here. Two methods were used to extract the g-factors in PbTe: The Zeeman splitting of quantized levels in a quantum dot (QD, Ref. \cite{Zonglin_Anisotropy}) and the quantized plateaus in a quantum point contact (QPC, Ref. \cite{Wenyu_Disorder, Yuhao_degeneracy}). Both methods yielded a g-factor below 20, which is significantly smaller than the values observed in Fig. 2. The primary structural difference in our device is the presence of the Pb film. We thus attribute this g-factor enhancement to the orbital effect induced by the superconducting film. Notably, utilizing orbital contributions (either from the nanowire or the superconductor) to enhance the g-factor and lower the critical field of zero modes has been proposed by a theory study \cite{Wimmer2017Orbital} and realized by experiments \cite{full_shell}.

\begin{figure*}[htb]
\includegraphics[width=\textwidth]{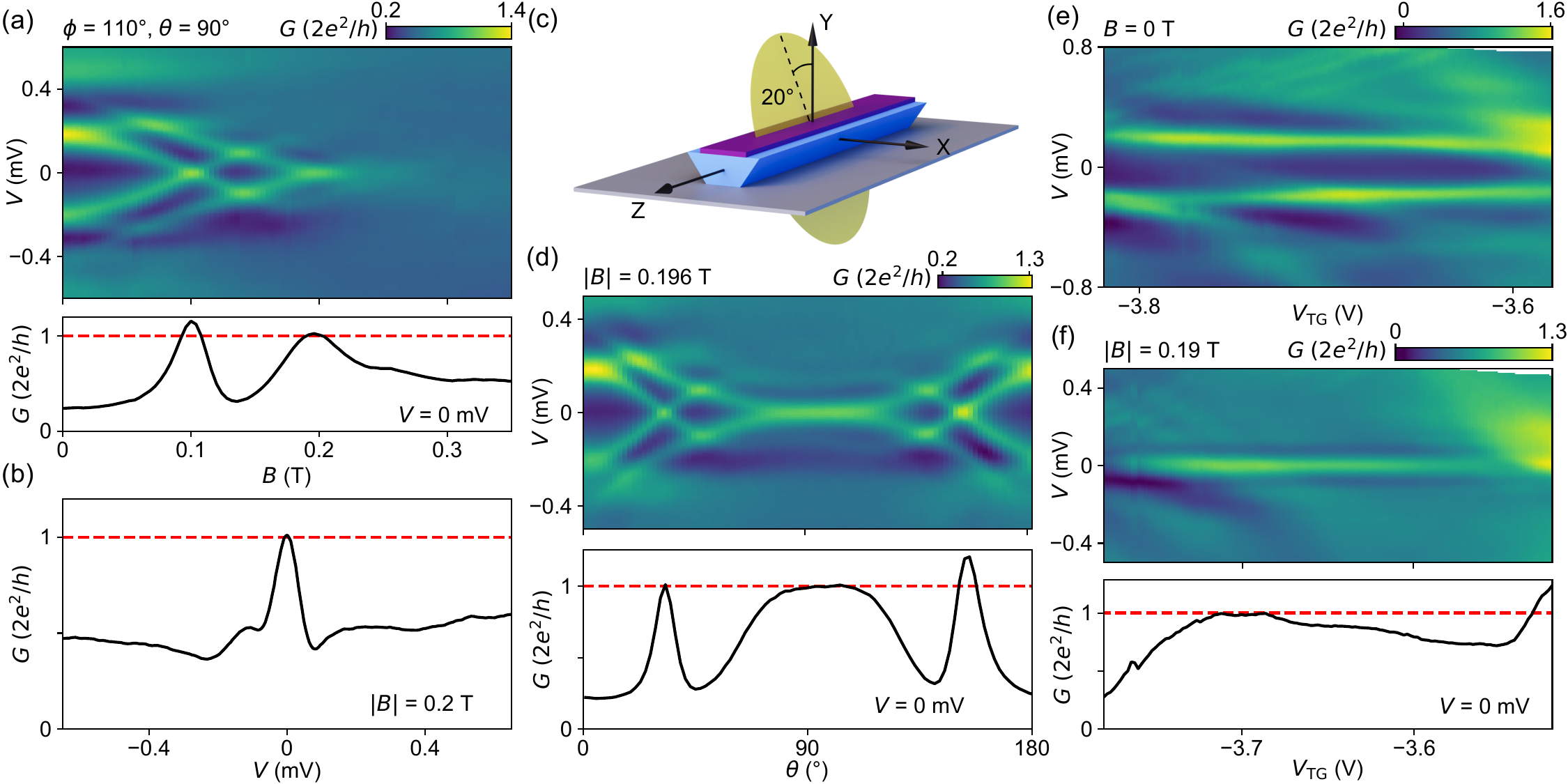}
\centering
\caption{(a) $B$ scan at $V_{\text{TG}}$ = -3.714 V, $V_{\text{SG}}$ = -3.4631 V. $\theta$ = 90$^{\circ}$, $\phi$ = 110$^{\circ}$. Lower panel, zero-bias line cut. (b) A line cut from (a) at 0.2 T.  (c-d) $B$ rotation by fixing $\phi$ = 110$^{\circ}$ and rotating $\theta$. The rotating plane is sketched in (c). Lower panel of (d) shows the zero-bias line cut. (e) $V_{\text{TG}}$ scan at zero field. $V_{\text{SG}}$ = -3.4631 V. (f) $V_{\text{TG}}$ scan at 0.19 T ($\theta$ = 90$^{\circ}$, $\phi$ = 110$^{\circ}$). $V_{\text{SG}}$ = -3.44 V. Lower panel, zero-bias line cut. }
\label{fig2}
\end{figure*}

We next oriented $B$ to deviate from the $y$ axis by 20$^{\circ}$ in the $xy$ plane ($\theta$ = 90$^{\circ}$, $\phi$ = 110$^{\circ}$). This direction corresponds to the ``sweet-spot'' angle that will be discussed later. Figure 2(c) presents the $B$ scan along this direction. The extracted g-factors are 75 for the inner subgap states and 70 for the outer ones. The inner states form a sharp crossing at zero energy, while the outer states form a zero-bias peak (ZBP) that persists over a sizable $B$ range. To better visualize this ZBP, Fig. 2(d) displays the normalized conductance: The maximum and minimum conductance values at each $B$ are scaled to 1 and 0, respectively. The dashed box highlights the ZBP, which persists from 0.175 T to 0.29 T, see Fig. 2(e) for the waterfall plot. This field range of 0.115 T corresponds to a Zeeman energy of 0.24 meV, comparable to values observed in InAs-Al nanowires \cite{Deng2016, WangZhaoyu, MS_2023}. A ZBP remaining robust (i.e. non-split) in $B$ and gate scans is a necessary but insufficient signature of Majorana zero modes. The gate scan of this device at zero field (Fig. S1) reveals QD behavior, indicating a significant amount of disorder. We thus cannot rule out a disorder origin \cite{Patrick_Lee_disorder_2012, Prada2012, Loss2013ZBP, Silvano2014, GoodBadUgly, DasSarma_estimate, Tudor2021Disorder, Loss_Andreev_band}. Future work should focus on reducing disorder in these devices to verify whether the robust ZBP remains.

Having established the large g-factor for $B$ near the out-of-plane direction, we next investigate its angle dependence. Figure 3(a) shows $B$ scans along different directions in the $xy$ plane: $\theta$ =90$^{\circ}$, $\phi$ (from left to right) corresponds to 0$^{\circ}$ (the $x$ axis), 30$^{\circ}$, 55$^{\circ}$, 75$^{\circ}$, and 90$^{\circ}$ (the $y$ axis). Figure 3(b) provides an illustration of these directions. The field at which the inner subgap states cross zero gradually increases from 0.1 T for the 90$^{\circ}$ case to 0.56 T for the 0$^{\circ}$ case. Correspondingly, the g-factor decreases from 68 to 15. This decrease is likely due to a reduction of the orbital contribution as $B$ is rotated towards the $x$ axis. When $B$ is parallel to the Pb film (the $x$ axis), the orbital contribution is negligible. And the observed g-factor of 15 is consistent with the bare PbTe case in Table \Romannum{1}. Figure 3(c) shows the $B$ scan along the nanowire axis, revealing a g-factor of 11. This value is close to that of the $x$ axis case, since both configurations are parallel to the Pb film.

Figures 3(d-f) show the continuous evolution of the subgap states as $B$ is rotated within the three planes. The upper panels illustrate these planes (highlighted in yellow). The field magnitude $|B|$ is fixed at 0.19 T, a value where the robust ZBP emerges in Fig. 2. In the $xy$ plane, the axis of symmetry is not the $y$ axis, but is deviated by 20$^{\circ}$ ($\phi$ = 110$^{\circ}$), as indicated by the orange bar in Fig. 3(d). We identify this angle ($\phi$ = 110$^{\circ}$, $\theta$ = 90$^{\circ}$) as the ``sweet-spot'' direction. The ZBP remains robust upon angle rotation (persisting over 46$^{\circ}$). The rotation measurement in Fig. 3(d) is equivalent to a line cut at 0.19 T across all the panels of Fig. 3(a). Thus the ZBP and its splitting during rotation is related to the g-factor anisotropy at different angles \cite{Silvano2014}. The mechanism of this anisotropy is closely linked to orbital effects of the Pb film.  For the other two planes in Figs. 3(e-f), the symmetric axes align with the $y$ and $x$ axes, respectively. Since the $xz$ plane is parallel to the Pb film, the g-factor is not enhanced, and no ZBP is observed at this field in Fig. 3(f).

We next apply $B$ along the sweet-spot direction. With an adjusted gate setting compared to Fig. 2, the ZBP height reaches $2e^2/h$, see Figs. 4(a-b). The lower panel shows its zero-bias line cut. The inner subgap states form a sharp crossing around 0.1 T with height exceeding $2e^2/h$. The outer states form a ZBP persisting over a sizable $B$ range, i.e. from 0.18 T to 0.28 T (see Fig. S3(a) for the normalized color plot). Although its maximum height reaches $2e^2/h$, the zero-bias line cut (lower panel of Fig. 4(a)) shows that the peak height does not exhibit a plateau upon sweeping $B$. The ZBP is thus not quantized. Recent theoretical simulations show that disorder could lead to accidental ZBPs at $2e^2/h$ \cite{GoodBadUgly}, which serves as a possible explanation for our observation.

We then fix $|B|$ at 0.196 T, $\phi$ at 110$^{\circ}$, and scan $\theta$. See Fig. 4(c) for the illustration of this plane. The $2e^2/h$-ZBP exhibits a plateau-like feature in Fig. 4(d), persisting from 79$^{\circ}$ to 115$^{\circ}$.  Figures 4(e-f) show the dependence on $V_{\text{TG}}$ near this gate setting at 0 T and 0.19 T, respectively. The zero-field scan does not reveal robust ZBPs. At 0.19 T, the ZBP persists over a sizable $V_{\text{TG}}$ range. Figure S1(c) shows a larger $V_{\text{TG}}$ range of this robust ZBP. The zero-bias line cut in the lower panel of Fig. 4(f) indicates that the ZBP height fluctuates near $2e^2/h$. The SG scan (Fig. S3(c)) reveals similar behavior. Note that true ZBP quantization requires a plateau against variations in all parameters. The plateau-like feature in the angular scan (Fig. 4(d)) alone is insufficient for quantization.

The g-factor enhancement for out-of-plane $B$ is a generic observation for PbTe-Pb hybrid nanowires. In Fig. S4, we show results from a second device. This device has less disorder, and its subgap states behavior is similar to that of device A. The maximum g-factor can reach 83. We note that in PbTe-In hybrid nanowires, the g-factor can occasionally reach 45 \cite{PbTe_In}. However, the induced gap of PbTe-In is around 1 meV. The critical field of topological phase transition would be higher than 0.7 T, several times larger than the case here (below 0.2 T). Moreover, robust ZBPs have not been observed in PbTe-In hybrids, implying the necessity of a large g-factor as well as a low critical field.

In conclusion, we have demonstrated large g-factors in PbTe-Pb hybrid nanowires. The g-factor can reach 75-83 for a nearly out-of-plane field, several times larger than that in PbTe nanowires with the same orientation. We attribute this enhancement to the orbital effects from the superconducting Pb film, which is corroborated by the anisotropy in $B$ rotation. This enhancement greatly reduces the magnetic field for the topological phase transition from above 1 T to below 0.2 T. We also observe zero-bias peaks at low fields along the direction of the maximum g-factor. These peaks remain robust (non-split) in gate and $B$ scans. Robust ZBPs are suggestive but insufficient evidence of Majoranas since disorder cannot be ruled out as an origin. Future studies could focus on further optimizing the nanowire quality --- such as fabricating cleaner devices with fewer Andreev bound states --- and implementing three-terminal end-to-end correlation measurements, which could provide stronger signatures of Majorana modes.  

\section{Acknowledgment} 

This work is supported by National Natural Science Foundation of China (Grant Nos. 92565302 and 12504191) and Quantum Science and Technology-National Science and Technology Major Project  (2021ZD0302400). W. S. acknowledges the Postdoctoral Fellowship Program and China Postdoctoral Science Foundation (Grant No. BX20250167 and No. 2025M783399). S.Y. acknowledges the China Postdoctoral Science Foundation (Grant No. 2024M751610) and Postdoctoral Fellowship Program of China Postdoctoral Science Foundation (Grant No. GZC20231368).

\section{Data availability}

Raw data and processing codes within this paper are available at https://doi.org/10.5281/zenodo.19183718

\bibliography{mybibfile}

\begin{thebibliography}{42}%
\makeatletter
\providecommand \@ifxundefined [1]{%
 \@ifx{#1\undefined}
}%
\providecommand \@ifnum [1]{%
 \ifnum #1\expandafter \@firstoftwo
 \else \expandafter \@secondoftwo
 \fi
}%
\providecommand \@ifx [1]{%
 \ifx #1\expandafter \@firstoftwo
 \else \expandafter \@secondoftwo
 \fi
}%
\providecommand \natexlab [1]{#1}%
\providecommand \enquote  [1]{``#1''}%
\providecommand \bibnamefont  [1]{#1}%
\providecommand \bibfnamefont [1]{#1}%
\providecommand \citenamefont [1]{#1}%
\providecommand \href@noop [0]{\@secondoftwo}%
\providecommand \href [0]{\begingroup \@sanitize@url \@href}%
\providecommand \@href[1]{\@@startlink{#1}\@@href}%
\providecommand \@@href[1]{\endgroup#1\@@endlink}%
\providecommand \@sanitize@url [0]{\catcode `\\12\catcode `\$12\catcode
  `\&12\catcode `\#12\catcode `\^12\catcode `\_12\catcode `\%12\relax}%
\providecommand \@@startlink[1]{}%
\providecommand \@@endlink[0]{}%
\providecommand \url  [0]{\begingroup\@sanitize@url \@url }%
\providecommand \@url [1]{\endgroup\@href {#1}{\urlprefix }}%
\providecommand \urlprefix  [0]{URL }%
\providecommand \Eprint [0]{\href }%
\providecommand \doibase [0]{https://doi.org/}%
\providecommand \selectlanguage [0]{\@gobble}%
\providecommand \bibinfo  [0]{\@secondoftwo}%
\providecommand \bibfield  [0]{\@secondoftwo}%
\providecommand \translation [1]{[#1]}%
\providecommand \BibitemOpen [0]{}%
\providecommand \bibitemStop [0]{}%
\providecommand \bibitemNoStop [0]{.\EOS\space}%
\providecommand \EOS [0]{\spacefactor3000\relax}%
\providecommand \BibitemShut  [1]{\csname bibitem#1\endcsname}%
\let\auto@bib@innerbib\@empty
\bibitem [{\citenamefont {Lutchyn}\ \emph {et~al.}(2010)\citenamefont
  {Lutchyn}, \citenamefont {Sau},\ and\ \citenamefont
  {Das~Sarma}}]{Lutchyn2010}%
  \BibitemOpen
  \bibfield  {author} {\bibinfo {author} {\bibfnamefont {R.~M.}\ \bibnamefont
  {Lutchyn}}, \bibinfo {author} {\bibfnamefont {J.~D.}\ \bibnamefont {Sau}},\
  and\ \bibinfo {author} {\bibfnamefont {S.}~\bibnamefont {Das~Sarma}},\
  }\bibfield  {title} {\bibinfo {title} {Majorana fermions and a topological
  phase transition in semiconductor-superconductor heterostructures},\ }\href
  {https://doi.org/10.1103/PhysRevLett.105.077001} {\bibfield  {journal}
  {\bibinfo  {journal} {Phys. Rev. Lett.}\ }\textbf {\bibinfo {volume} {105}},\
  \bibinfo {pages} {077001} (\bibinfo {year} {2010})}\BibitemShut {NoStop}%
\bibitem [{\citenamefont {Oreg}\ \emph {et~al.}(2010)\citenamefont {Oreg},
  \citenamefont {Refael},\ and\ \citenamefont {von Oppen}}]{Oreg2010}%
  \BibitemOpen
  \bibfield  {author} {\bibinfo {author} {\bibfnamefont {Y.}~\bibnamefont
  {Oreg}}, \bibinfo {author} {\bibfnamefont {G.}~\bibnamefont {Refael}},\ and\
  \bibinfo {author} {\bibfnamefont {F.}~\bibnamefont {von Oppen}},\ }\bibfield
  {title} {\bibinfo {title} {Helical liquids and {M}ajorana bound states in
  quantum wires},\ }\href {https://doi.org/10.1103/PhysRevLett.105.177002}
  {\bibfield  {journal} {\bibinfo  {journal} {Phys. Rev. Lett.}\ }\textbf
  {\bibinfo {volume} {105}},\ \bibinfo {pages} {177002} (\bibinfo {year}
  {2010})}\BibitemShut {NoStop}%
\bibitem [{\citenamefont {Mourik}\ \emph {et~al.}(2012)\citenamefont {Mourik},
  \citenamefont {Zuo}, \citenamefont {Frolov}, \citenamefont {Plissard},
  \citenamefont {Bakkers},\ and\ \citenamefont {Kouwenhoven}}]{Mourik}%
  \BibitemOpen
  \bibfield  {author} {\bibinfo {author} {\bibfnamefont {V.}~\bibnamefont
  {Mourik}}, \bibinfo {author} {\bibfnamefont {K.}~\bibnamefont {Zuo}},
  \bibinfo {author} {\bibfnamefont {S.~M.}\ \bibnamefont {Frolov}}, \bibinfo
  {author} {\bibfnamefont {S.}~\bibnamefont {Plissard}}, \bibinfo {author}
  {\bibfnamefont {E.~P.}\ \bibnamefont {Bakkers}},\ and\ \bibinfo {author}
  {\bibfnamefont {L.~P.}\ \bibnamefont {Kouwenhoven}},\ }\bibfield  {title}
  {\bibinfo {title} {Signatures of {M}ajorana fermions in hybrid
  superconductor-semiconductor nanowire devices},\ }\href
  {https://doi.org/10.1126/science.1222360} {\bibfield  {journal} {\bibinfo
  {journal} {Science}\ }\textbf {\bibinfo {volume} {336}},\ \bibinfo {pages}
  {1003} (\bibinfo {year} {2012})}\BibitemShut {NoStop}%
\bibitem [{\citenamefont {Deng}\ \emph {et~al.}(2016)\citenamefont {Deng},
  \citenamefont {Vaitiek{\.e}nas}, \citenamefont {Hansen}, \citenamefont
  {Danon}, \citenamefont {Leijnse}, \citenamefont {Flensberg}, \citenamefont
  {Nyg{\aa}rd}, \citenamefont {Krogstrup},\ and\ \citenamefont
  {Marcus}}]{Deng2016}%
  \BibitemOpen
  \bibfield  {author} {\bibinfo {author} {\bibfnamefont {M.}~\bibnamefont
  {Deng}}, \bibinfo {author} {\bibfnamefont {S.}~\bibnamefont
  {Vaitiek{\.e}nas}}, \bibinfo {author} {\bibfnamefont {E.~B.}\ \bibnamefont
  {Hansen}}, \bibinfo {author} {\bibfnamefont {J.}~\bibnamefont {Danon}},
  \bibinfo {author} {\bibfnamefont {M.}~\bibnamefont {Leijnse}}, \bibinfo
  {author} {\bibfnamefont {K.}~\bibnamefont {Flensberg}}, \bibinfo {author}
  {\bibfnamefont {J.}~\bibnamefont {Nyg{\aa}rd}}, \bibinfo {author}
  {\bibfnamefont {P.}~\bibnamefont {Krogstrup}},\ and\ \bibinfo {author}
  {\bibfnamefont {C.~M.}\ \bibnamefont {Marcus}},\ }\bibfield  {title}
  {\bibinfo {title} {Majorana bound state in a coupled quantum-dot
  hybrid-nanowire system},\ }\href {https://doi.org/10.1126/science.aaf3961}
  {\bibfield  {journal} {\bibinfo  {journal} {Science}\ }\textbf {\bibinfo
  {volume} {354}},\ \bibinfo {pages} {1557} (\bibinfo {year}
  {2016})}\BibitemShut {NoStop}%
\bibitem [{\citenamefont {G{\"u}l}\ \emph {et~al.}(2018)\citenamefont
  {G{\"u}l}, \citenamefont {Zhang}, \citenamefont {Bommer}, \citenamefont
  {de~Moor}, \citenamefont {Car}, \citenamefont {Plissard}, \citenamefont
  {Bakkers}, \citenamefont {Geresdi}, \citenamefont {Watanabe}, \citenamefont
  {Taniguchi} \emph {et~al.}}]{Gul2018}%
  \BibitemOpen
  \bibfield  {author} {\bibinfo {author} {\bibfnamefont {{\"O}.}~\bibnamefont
  {G{\"u}l}}, \bibinfo {author} {\bibfnamefont {H.}~\bibnamefont {Zhang}},
  \bibinfo {author} {\bibfnamefont {J.~D.}\ \bibnamefont {Bommer}}, \bibinfo
  {author} {\bibfnamefont {M.~W.}\ \bibnamefont {de~Moor}}, \bibinfo {author}
  {\bibfnamefont {D.}~\bibnamefont {Car}}, \bibinfo {author} {\bibfnamefont
  {S.~R.}\ \bibnamefont {Plissard}}, \bibinfo {author} {\bibfnamefont {E.~P.}\
  \bibnamefont {Bakkers}}, \bibinfo {author} {\bibfnamefont {A.}~\bibnamefont
  {Geresdi}}, \bibinfo {author} {\bibfnamefont {K.}~\bibnamefont {Watanabe}},
  \bibinfo {author} {\bibfnamefont {T.}~\bibnamefont {Taniguchi}}, \emph
  {et~al.},\ }\bibfield  {title} {\bibinfo {title} {Ballistic {M}ajorana
  nanowire devices},\ }\href {https://doi.org/10.1038/s41565-017-0032-8}
  {\bibfield  {journal} {\bibinfo  {journal} {Nature Nanotechnology}\ }\textbf
  {\bibinfo {volume} {13}},\ \bibinfo {pages} {192} (\bibinfo {year}
  {2018})}\BibitemShut {NoStop}%
\bibitem [{\citenamefont {Song}\ \emph {et~al.}(2022)\citenamefont {Song},
  \citenamefont {Zhang}, \citenamefont {Pan}, \citenamefont {Liu},
  \citenamefont {Wang}, \citenamefont {Cao}, \citenamefont {Liu}, \citenamefont
  {Wen}, \citenamefont {Liao}, \citenamefont {Zhuo} \emph {et~al.}}]{Song2022}%
  \BibitemOpen
  \bibfield  {author} {\bibinfo {author} {\bibfnamefont {H.}~\bibnamefont
  {Song}}, \bibinfo {author} {\bibfnamefont {Z.}~\bibnamefont {Zhang}},
  \bibinfo {author} {\bibfnamefont {D.}~\bibnamefont {Pan}}, \bibinfo {author}
  {\bibfnamefont {D.}~\bibnamefont {Liu}}, \bibinfo {author} {\bibfnamefont
  {Z.}~\bibnamefont {Wang}}, \bibinfo {author} {\bibfnamefont {Z.}~\bibnamefont
  {Cao}}, \bibinfo {author} {\bibfnamefont {L.}~\bibnamefont {Liu}}, \bibinfo
  {author} {\bibfnamefont {L.}~\bibnamefont {Wen}}, \bibinfo {author}
  {\bibfnamefont {D.}~\bibnamefont {Liao}}, \bibinfo {author} {\bibfnamefont
  {R.}~\bibnamefont {Zhuo}}, \emph {et~al.},\ }\bibfield  {title} {\bibinfo
  {title} {Large zero bias peaks and dips in a four-terminal thin {I}n{A}s-{A}l
  nanowire device},\ }\href {https://doi.org/10.1103/PhysRevResearch.4.033235}
  {\bibfield  {journal} {\bibinfo  {journal} {Phys. Rev. Research}\ }\textbf
  {\bibinfo {volume} {4}},\ \bibinfo {pages} {033235} (\bibinfo {year}
  {2022})}\BibitemShut {NoStop}%
\bibitem [{\citenamefont {Wang}\ \emph {et~al.}(2022)\citenamefont {Wang},
  \citenamefont {Song}, \citenamefont {Pan}, \citenamefont {Zhang},
  \citenamefont {Miao}, \citenamefont {Li}, \citenamefont {Cao}, \citenamefont
  {Zhang}, \citenamefont {Liu}, \citenamefont {Wen} \emph
  {et~al.}}]{WangZhaoyu}%
  \BibitemOpen
  \bibfield  {author} {\bibinfo {author} {\bibfnamefont {Z.}~\bibnamefont
  {Wang}}, \bibinfo {author} {\bibfnamefont {H.}~\bibnamefont {Song}}, \bibinfo
  {author} {\bibfnamefont {D.}~\bibnamefont {Pan}}, \bibinfo {author}
  {\bibfnamefont {Z.}~\bibnamefont {Zhang}}, \bibinfo {author} {\bibfnamefont
  {W.}~\bibnamefont {Miao}}, \bibinfo {author} {\bibfnamefont {R.}~\bibnamefont
  {Li}}, \bibinfo {author} {\bibfnamefont {Z.}~\bibnamefont {Cao}}, \bibinfo
  {author} {\bibfnamefont {G.}~\bibnamefont {Zhang}}, \bibinfo {author}
  {\bibfnamefont {L.}~\bibnamefont {Liu}}, \bibinfo {author} {\bibfnamefont
  {L.}~\bibnamefont {Wen}}, \emph {et~al.},\ }\bibfield  {title} {\bibinfo
  {title} {Plateau regions for zero-bias peaks within 5$\%$ of the quantized
  conductance value $2{e}^{2}/h$},\ }\href
  {https://doi.org/10.1103/PhysRevLett.129.167702} {\bibfield  {journal}
  {\bibinfo  {journal} {Phys. Rev. Lett.}\ }\textbf {\bibinfo {volume} {129}},\
  \bibinfo {pages} {167702} (\bibinfo {year} {2022})}\BibitemShut {NoStop}%
\bibitem [{\citenamefont {Dvir}\ \emph {et~al.}(2023)\citenamefont {Dvir},
  \citenamefont {Wang}, \citenamefont {van Loo}, \citenamefont {Liu},
  \citenamefont {Mazur}, \citenamefont {Bordin}, \citenamefont {Haaf},
  \citenamefont {Wang}, \citenamefont {Driel}, \citenamefont {Zatelli} \emph
  {et~al.}}]{Delft_Kitaev}%
  \BibitemOpen
  \bibfield  {author} {\bibinfo {author} {\bibfnamefont {T.}~\bibnamefont
  {Dvir}}, \bibinfo {author} {\bibfnamefont {G.}~\bibnamefont {Wang}}, \bibinfo
  {author} {\bibfnamefont {N.}~\bibnamefont {van Loo}}, \bibinfo {author}
  {\bibfnamefont {C.-X.}\ \bibnamefont {Liu}}, \bibinfo {author} {\bibfnamefont
  {G.}~\bibnamefont {Mazur}}, \bibinfo {author} {\bibfnamefont
  {A.}~\bibnamefont {Bordin}}, \bibinfo {author} {\bibfnamefont
  {S.}~\bibnamefont {Haaf}}, \bibinfo {author} {\bibfnamefont {J.-Y.}\
  \bibnamefont {Wang}}, \bibinfo {author} {\bibfnamefont {D.}~\bibnamefont
  {Driel}}, \bibinfo {author} {\bibfnamefont {F.}~\bibnamefont {Zatelli}},
  \emph {et~al.},\ }\bibfield  {title} {\bibinfo {title} {Realization of a
  minimal {Kitaev} chain in coupled quantum dots},\ }\href
  {https://doi.org/10.1038/s41586-022-05585-1} {\bibfield  {journal} {\bibinfo
  {journal} {Nature}\ }\textbf {\bibinfo {volume} {614}},\ \bibinfo {pages}
  {445} (\bibinfo {year} {2023})}\BibitemShut {NoStop}%
\bibitem [{\citenamefont {Aghaee}\ \emph {et~al.}(2023)\citenamefont {Aghaee},
  \citenamefont {Akkala}, \citenamefont {Alam}, \citenamefont {Ali},
  \citenamefont {Alcaraz~Ramirez}, \citenamefont {Andrzejczuk}, \citenamefont
  {Antipov}, \citenamefont {Aseev}, \citenamefont {Astafev}, \citenamefont
  {Bauer} \emph {et~al.}}]{MS_2023}%
  \BibitemOpen
  \bibfield  {author} {\bibinfo {author} {\bibfnamefont {M.}~\bibnamefont
  {Aghaee}}, \bibinfo {author} {\bibfnamefont {A.}~\bibnamefont {Akkala}},
  \bibinfo {author} {\bibfnamefont {Z.}~\bibnamefont {Alam}}, \bibinfo {author}
  {\bibfnamefont {R.}~\bibnamefont {Ali}}, \bibinfo {author} {\bibfnamefont
  {A.}~\bibnamefont {Alcaraz~Ramirez}}, \bibinfo {author} {\bibfnamefont
  {M.}~\bibnamefont {Andrzejczuk}}, \bibinfo {author} {\bibfnamefont {A.~E.}\
  \bibnamefont {Antipov}}, \bibinfo {author} {\bibfnamefont {P.}~\bibnamefont
  {Aseev}}, \bibinfo {author} {\bibfnamefont {M.}~\bibnamefont {Astafev}},
  \bibinfo {author} {\bibfnamefont {B.}~\bibnamefont {Bauer}}, \emph {et~al.},\
  }\bibfield  {title} {\bibinfo {title} {{InAs-Al} hybrid devices passing the
  topological gap protocol},\ }\href
  {https://doi.org/10.1103/PhysRevB.107.245423} {\bibfield  {journal} {\bibinfo
   {journal} {Phys. Rev. B}\ }\textbf {\bibinfo {volume} {107}},\ \bibinfo
  {pages} {245423} (\bibinfo {year} {2023})}\BibitemShut {NoStop}%
\bibitem [{\citenamefont {Aghaee}\ \emph {et~al.}(2025)\citenamefont {Aghaee},
  \citenamefont {Ramirez}, \citenamefont {Alam}, \citenamefont {Ali},
  \citenamefont {Andrzejczuk}, \citenamefont {Antipov}, \citenamefont
  {Astafev}, \citenamefont {Barzegar}, \citenamefont {Bauer} \emph
  {et~al.}}]{MS_2025}%
  \BibitemOpen
  \bibfield  {author} {\bibinfo {author} {\bibfnamefont {M.}~\bibnamefont
  {Aghaee}}, \bibinfo {author} {\bibfnamefont {A.}~\bibnamefont {Ramirez}},
  \bibinfo {author} {\bibfnamefont {Z.}~\bibnamefont {Alam}}, \bibinfo {author}
  {\bibfnamefont {R.}~\bibnamefont {Ali}}, \bibinfo {author} {\bibfnamefont
  {M.}~\bibnamefont {Andrzejczuk}}, \bibinfo {author} {\bibfnamefont
  {A.}~\bibnamefont {Antipov}}, \bibinfo {author} {\bibfnamefont
  {M.}~\bibnamefont {Astafev}}, \bibinfo {author} {\bibfnamefont
  {A.}~\bibnamefont {Barzegar}}, \bibinfo {author} {\bibfnamefont
  {B.}~\bibnamefont {Bauer}}, \emph {et~al.},\ }\bibfield  {title} {\bibinfo
  {title} {Interferometric single-shot parity measurement in {InAs–Al} hybrid
  devices},\ }\href {https://doi.org/10.1038/s41586-024-08445-2} {\bibfield
  {journal} {\bibinfo  {journal} {Nature}\ }\textbf {\bibinfo {volume} {638}},\
  \bibinfo {pages} {651} (\bibinfo {year} {2025})}\BibitemShut {NoStop}%
\bibitem [{\citenamefont {Zhang}\ \emph {et~al.}(2019)\citenamefont {Zhang},
  \citenamefont {Liu}, \citenamefont {Wimmer},\ and\ \citenamefont
  {Kouwenhoven}}]{NextSteps}%
  \BibitemOpen
  \bibfield  {author} {\bibinfo {author} {\bibfnamefont {H.}~\bibnamefont
  {Zhang}}, \bibinfo {author} {\bibfnamefont {D.~E.}\ \bibnamefont {Liu}},
  \bibinfo {author} {\bibfnamefont {M.}~\bibnamefont {Wimmer}},\ and\ \bibinfo
  {author} {\bibfnamefont {L.~P.}\ \bibnamefont {Kouwenhoven}},\ }\bibfield
  {title} {\bibinfo {title} {Next steps of quantum transport in {M}ajorana
  nanowire devices},\ }\href {https://doi.org/10.1038/s41467-019-13133-1}
  {\bibfield  {journal} {\bibinfo  {journal} {Nature Communications}\ }\textbf
  {\bibinfo {volume} {10}},\ \bibinfo {pages} {5128} (\bibinfo {year}
  {2019})}\BibitemShut {NoStop}%
\bibitem [{\citenamefont {Prada}\ \emph {et~al.}(2020)\citenamefont {Prada},
  \citenamefont {San-Jose}, \citenamefont {de~Moor}, \citenamefont {Geresdi},
  \citenamefont {Lee}, \citenamefont {Klinovaja}, \citenamefont {Loss},
  \citenamefont {Nyg{\aa}rd}, \citenamefont {Aguado},\ and\ \citenamefont
  {Kouwenhoven}}]{Prada2020}%
  \BibitemOpen
  \bibfield  {author} {\bibinfo {author} {\bibfnamefont {E.}~\bibnamefont
  {Prada}}, \bibinfo {author} {\bibfnamefont {P.}~\bibnamefont {San-Jose}},
  \bibinfo {author} {\bibfnamefont {M.~W.}\ \bibnamefont {de~Moor}}, \bibinfo
  {author} {\bibfnamefont {A.}~\bibnamefont {Geresdi}}, \bibinfo {author}
  {\bibfnamefont {E.~J.}\ \bibnamefont {Lee}}, \bibinfo {author} {\bibfnamefont
  {J.}~\bibnamefont {Klinovaja}}, \bibinfo {author} {\bibfnamefont
  {D.}~\bibnamefont {Loss}}, \bibinfo {author} {\bibfnamefont {J.}~\bibnamefont
  {Nyg{\aa}rd}}, \bibinfo {author} {\bibfnamefont {R.}~\bibnamefont {Aguado}},\
  and\ \bibinfo {author} {\bibfnamefont {L.~P.}\ \bibnamefont {Kouwenhoven}},\
  }\bibfield  {title} {\bibinfo {title} {From {A}ndreev to {M}ajorana bound
  states in hybrid superconductor--semiconductor nanowires},\ }\href
  {https://doi.org/10.1038/s42254-020-0228-y} {\bibfield  {journal} {\bibinfo
  {journal} {Nature Reviews Physics}\ }\textbf {\bibinfo {volume} {2}},\
  \bibinfo {pages} {575} (\bibinfo {year} {2020})}\BibitemShut {NoStop}%
\bibitem [{\citenamefont {Kouwenhoven}(2025)}]{Leo_perspective}%
  \BibitemOpen
  \bibfield  {author} {\bibinfo {author} {\bibfnamefont {L.}~\bibnamefont
  {Kouwenhoven}},\ }\bibfield  {title} {\bibinfo {title} {Perspective on
  {M}ajorana bound-states in hybrid superconductor-semiconductor nanowires},\
  }\href {https://doi.org/10.1142/S0217984925400020} {\bibfield  {journal}
  {\bibinfo  {journal} {Modern Physics Letters B}\ }\textbf {\bibinfo {volume}
  {39}},\ \bibinfo {pages} {2540002} (\bibinfo {year} {2025})}\BibitemShut
  {NoStop}%
\bibitem [{\citenamefont {Kammhuber}\ \emph {et~al.}(2016)\citenamefont
  {Kammhuber}, \citenamefont {Cassidy}, \citenamefont {Zhang}, \citenamefont
  {G{\"u}l}, \citenamefont {Pei}, \citenamefont {de~Moor}, \citenamefont
  {Nijholt}, \citenamefont {Watanabe}, \citenamefont {Taniguchi}, \citenamefont
  {Car} \emph {et~al.}}]{Kammhuber2016}%
  \BibitemOpen
  \bibfield  {author} {\bibinfo {author} {\bibfnamefont {J.}~\bibnamefont
  {Kammhuber}}, \bibinfo {author} {\bibfnamefont {M.~C.}\ \bibnamefont
  {Cassidy}}, \bibinfo {author} {\bibfnamefont {H.}~\bibnamefont {Zhang}},
  \bibinfo {author} {\bibfnamefont {{\"O}.}~\bibnamefont {G{\"u}l}}, \bibinfo
  {author} {\bibfnamefont {F.}~\bibnamefont {Pei}}, \bibinfo {author}
  {\bibfnamefont {M.~W.}\ \bibnamefont {de~Moor}}, \bibinfo {author}
  {\bibfnamefont {B.}~\bibnamefont {Nijholt}}, \bibinfo {author} {\bibfnamefont
  {K.}~\bibnamefont {Watanabe}}, \bibinfo {author} {\bibfnamefont
  {T.}~\bibnamefont {Taniguchi}}, \bibinfo {author} {\bibfnamefont
  {D.}~\bibnamefont {Car}}, \emph {et~al.},\ }\bibfield  {title} {\bibinfo
  {title} {Conductance quantization at zero magnetic field in {I}n{S}b
  nanowires},\ }\href {https://doi.org/10.1021/acs.nanolett.6b00051} {\bibfield
   {journal} {\bibinfo  {journal} {Nano Letters}\ }\textbf {\bibinfo {volume}
  {16}},\ \bibinfo {pages} {3482} (\bibinfo {year} {2016})}\BibitemShut
  {NoStop}%
\bibitem [{\citenamefont {Reeg}\ \emph {et~al.}(2017)\citenamefont {Reeg},
  \citenamefont {Loss},\ and\ \citenamefont {Klinovaja}}]{Loss_metalization}%
  \BibitemOpen
  \bibfield  {author} {\bibinfo {author} {\bibfnamefont {C.}~\bibnamefont
  {Reeg}}, \bibinfo {author} {\bibfnamefont {D.}~\bibnamefont {Loss}},\ and\
  \bibinfo {author} {\bibfnamefont {J.}~\bibnamefont {Klinovaja}},\ }\bibfield
  {title} {\bibinfo {title} {Finite-size effects in a nanowire strongly coupled
  to a thin superconducting shell},\ }\href
  {https://doi.org/10.1103/PhysRevB.96.125426} {\bibfield  {journal} {\bibinfo
  {journal} {Phys. Rev. B}\ }\textbf {\bibinfo {volume} {96}},\ \bibinfo
  {pages} {125426} (\bibinfo {year} {2017})}\BibitemShut {NoStop}%
\bibitem [{\citenamefont {Cao}\ \emph {et~al.}(2022)\citenamefont {Cao},
  \citenamefont {Liu}, \citenamefont {He}, \citenamefont {Liu}, \citenamefont
  {He},\ and\ \citenamefont {Zhang}}]{CaoZhanPbTe}%
  \BibitemOpen
  \bibfield  {author} {\bibinfo {author} {\bibfnamefont {Z.}~\bibnamefont
  {Cao}}, \bibinfo {author} {\bibfnamefont {D.~E.}\ \bibnamefont {Liu}},
  \bibinfo {author} {\bibfnamefont {W.-X.}\ \bibnamefont {He}}, \bibinfo
  {author} {\bibfnamefont {X.}~\bibnamefont {Liu}}, \bibinfo {author}
  {\bibfnamefont {K.}~\bibnamefont {He}},\ and\ \bibinfo {author}
  {\bibfnamefont {H.}~\bibnamefont {Zhang}},\ }\bibfield  {title} {\bibinfo
  {title} {Numerical study of {P}b{T}e-{P}b hybrid nanowires for engineering
  {M}ajorana zero modes},\ }\href {https://doi.org/10.1103/PhysRevB.105.085424}
  {\bibfield  {journal} {\bibinfo  {journal} {Phys. Rev. B}\ }\textbf {\bibinfo
  {volume} {105}},\ \bibinfo {pages} {085424} (\bibinfo {year}
  {2022})}\BibitemShut {NoStop}%
\bibitem [{\citenamefont {Jiang}\ \emph {et~al.}(2022)\citenamefont {Jiang},
  \citenamefont {Yang}, \citenamefont {Li}, \citenamefont {Song}, \citenamefont
  {Miao}, \citenamefont {Tong}, \citenamefont {Geng}, \citenamefont {Gao},
  \citenamefont {Li}, \citenamefont {Chen} \emph {et~al.}}]{Jiangyuying}%
  \BibitemOpen
  \bibfield  {author} {\bibinfo {author} {\bibfnamefont {Y.}~\bibnamefont
  {Jiang}}, \bibinfo {author} {\bibfnamefont {S.}~\bibnamefont {Yang}},
  \bibinfo {author} {\bibfnamefont {L.}~\bibnamefont {Li}}, \bibinfo {author}
  {\bibfnamefont {W.}~\bibnamefont {Song}}, \bibinfo {author} {\bibfnamefont
  {W.}~\bibnamefont {Miao}}, \bibinfo {author} {\bibfnamefont {B.}~\bibnamefont
  {Tong}}, \bibinfo {author} {\bibfnamefont {Z.}~\bibnamefont {Geng}}, \bibinfo
  {author} {\bibfnamefont {Y.}~\bibnamefont {Gao}}, \bibinfo {author}
  {\bibfnamefont {R.}~\bibnamefont {Li}}, \bibinfo {author} {\bibfnamefont
  {F.}~\bibnamefont {Chen}}, \emph {et~al.},\ }\bibfield  {title} {\bibinfo
  {title} {Selective area epitaxy of {P}b{T}e-{P}b hybrid nanowires on a
  lattice-matched substrate},\ }\href
  {https://doi.org/10.1103/PhysRevMaterials.6.034205} {\bibfield  {journal}
  {\bibinfo  {journal} {Phys. Rev. Materials}\ }\textbf {\bibinfo {volume}
  {6}},\ \bibinfo {pages} {034205} (\bibinfo {year} {2022})}\BibitemShut
  {NoStop}%
\bibitem [{\citenamefont {Jung}\ \emph {et~al.}(2022)\citenamefont {Jung},
  \citenamefont {Schellingerhout}, \citenamefont {Ritter}, \citenamefont {ten
  Kate}, \citenamefont {van~der Molen}, \citenamefont {de~Loijer},
  \citenamefont {Verheijen}, \citenamefont {Riel}, \citenamefont {Nichele},\
  and\ \citenamefont {Bakkers}}]{Erik_PbTe_SAG}%
  \BibitemOpen
  \bibfield  {author} {\bibinfo {author} {\bibfnamefont {J.}~\bibnamefont
  {Jung}}, \bibinfo {author} {\bibfnamefont {S.~G.}\ \bibnamefont
  {Schellingerhout}}, \bibinfo {author} {\bibfnamefont {M.~F.}\ \bibnamefont
  {Ritter}}, \bibinfo {author} {\bibfnamefont {S.~C.}\ \bibnamefont {ten
  Kate}}, \bibinfo {author} {\bibfnamefont {O.~A.}\ \bibnamefont {van~der
  Molen}}, \bibinfo {author} {\bibfnamefont {S.}~\bibnamefont {de~Loijer}},
  \bibinfo {author} {\bibfnamefont {M.~A.}\ \bibnamefont {Verheijen}}, \bibinfo
  {author} {\bibfnamefont {H.}~\bibnamefont {Riel}}, \bibinfo {author}
  {\bibfnamefont {F.}~\bibnamefont {Nichele}},\ and\ \bibinfo {author}
  {\bibfnamefont {E.~P.}\ \bibnamefont {Bakkers}},\ }\bibfield  {title}
  {\bibinfo {title} {Selective area growth of {P}b{T}e nanowire networks on
  {I}n{P}},\ }\href {https://doi.org/https://doi.org/10.1002/adfm.202208974}
  {\bibfield  {journal} {\bibinfo  {journal} {Advanced Functional Materials}\
  }\textbf {\bibinfo {volume} {32}},\ \bibinfo {pages} {2208974} (\bibinfo
  {year} {2022})}\BibitemShut {NoStop}%
\bibitem [{\citenamefont {Geng}\ \emph {et~al.}(2022)\citenamefont {Geng},
  \citenamefont {Zhang}, \citenamefont {Chen}, \citenamefont {Yang},
  \citenamefont {Jiang}, \citenamefont {Gao}, \citenamefont {Tong},
  \citenamefont {Song}, \citenamefont {Miao}, \citenamefont {Li} \emph
  {et~al.}}]{PbTe_AB}%
  \BibitemOpen
  \bibfield  {author} {\bibinfo {author} {\bibfnamefont {Z.}~\bibnamefont
  {Geng}}, \bibinfo {author} {\bibfnamefont {Z.}~\bibnamefont {Zhang}},
  \bibinfo {author} {\bibfnamefont {F.}~\bibnamefont {Chen}}, \bibinfo {author}
  {\bibfnamefont {S.}~\bibnamefont {Yang}}, \bibinfo {author} {\bibfnamefont
  {Y.}~\bibnamefont {Jiang}}, \bibinfo {author} {\bibfnamefont
  {Y.}~\bibnamefont {Gao}}, \bibinfo {author} {\bibfnamefont {B.}~\bibnamefont
  {Tong}}, \bibinfo {author} {\bibfnamefont {W.}~\bibnamefont {Song}}, \bibinfo
  {author} {\bibfnamefont {W.}~\bibnamefont {Miao}}, \bibinfo {author}
  {\bibfnamefont {R.}~\bibnamefont {Li}}, \emph {et~al.},\ }\bibfield  {title}
  {\bibinfo {title} {Observation of {A}haronov-{B}ohm effect in {P}b{T}e
  nanowire networks},\ }\href {https://doi.org/10.1103/PhysRevB.105.L241112}
  {\bibfield  {journal} {\bibinfo  {journal} {Phys. Rev. B}\ }\textbf {\bibinfo
  {volume} {105}},\ \bibinfo {pages} {L241112} (\bibinfo {year}
  {2022})}\BibitemShut {NoStop}%
\bibitem [{\citenamefont {ten Kate}\ \emph {et~al.}(2022)\citenamefont {ten
  Kate}, \citenamefont {Ritter}, \citenamefont {Fuhrer}, \citenamefont {Jung},
  \citenamefont {Schellingerhout}, \citenamefont {Bakkers}, \citenamefont
  {Riel},\ and\ \citenamefont {Nichele}}]{Fabrizio_PbTe}%
  \BibitemOpen
  \bibfield  {author} {\bibinfo {author} {\bibfnamefont {S.~C.}\ \bibnamefont
  {ten Kate}}, \bibinfo {author} {\bibfnamefont {M.~F.}\ \bibnamefont
  {Ritter}}, \bibinfo {author} {\bibfnamefont {A.}~\bibnamefont {Fuhrer}},
  \bibinfo {author} {\bibfnamefont {J.}~\bibnamefont {Jung}}, \bibinfo {author}
  {\bibfnamefont {S.~G.}\ \bibnamefont {Schellingerhout}}, \bibinfo {author}
  {\bibfnamefont {E.~P. A.~M.}\ \bibnamefont {Bakkers}}, \bibinfo {author}
  {\bibfnamefont {H.}~\bibnamefont {Riel}},\ and\ \bibinfo {author}
  {\bibfnamefont {F.}~\bibnamefont {Nichele}},\ }\bibfield  {title} {\bibinfo
  {title} {Small charging energies and g-factor anisotropy in {PbTe} quantum
  dots},\ }\href {https://doi.org/10.1021/acs.nanolett.2c01943} {\bibfield
  {journal} {\bibinfo  {journal} {Nano Letters}\ }\textbf {\bibinfo {volume}
  {22}},\ \bibinfo {pages} {7049} (\bibinfo {year} {2022})}\BibitemShut
  {NoStop}%
\bibitem [{\citenamefont {Zhang}\ \emph {et~al.}(2023)\citenamefont {Zhang},
  \citenamefont {Song}, \citenamefont {Gao}, \citenamefont {Wang},
  \citenamefont {Yu}, \citenamefont {Yang}, \citenamefont {Jiang},
  \citenamefont {Miao}, \citenamefont {Li}, \citenamefont {Chen} \emph
  {et~al.}}]{Zitong_JJ}%
  \BibitemOpen
  \bibfield  {author} {\bibinfo {author} {\bibfnamefont {Z.}~\bibnamefont
  {Zhang}}, \bibinfo {author} {\bibfnamefont {W.}~\bibnamefont {Song}},
  \bibinfo {author} {\bibfnamefont {Y.}~\bibnamefont {Gao}}, \bibinfo {author}
  {\bibfnamefont {Y.}~\bibnamefont {Wang}}, \bibinfo {author} {\bibfnamefont
  {Z.}~\bibnamefont {Yu}}, \bibinfo {author} {\bibfnamefont {S.}~\bibnamefont
  {Yang}}, \bibinfo {author} {\bibfnamefont {Y.}~\bibnamefont {Jiang}},
  \bibinfo {author} {\bibfnamefont {W.}~\bibnamefont {Miao}}, \bibinfo {author}
  {\bibfnamefont {R.}~\bibnamefont {Li}}, \bibinfo {author} {\bibfnamefont
  {F.}~\bibnamefont {Chen}}, \emph {et~al.},\ }\bibfield  {title} {\bibinfo
  {title} {Proximity effect in {PbTe-Pb} hybrid nanowire {J}osephson
  junctions},\ }\href {https://doi.org/10.1103/PhysRevMaterials.7.086201}
  {\bibfield  {journal} {\bibinfo  {journal} {Phys. Rev. Mater.}\ }\textbf
  {\bibinfo {volume} {7}},\ \bibinfo {pages} {086201} (\bibinfo {year}
  {2023})}\BibitemShut {NoStop}%
\bibitem [{\citenamefont {Song}\ \emph {et~al.}(2023)\citenamefont {Song},
  \citenamefont {Wang}, \citenamefont {Miao}, \citenamefont {Yu}, \citenamefont
  {Gao}, \citenamefont {Li}, \citenamefont {Yang}, \citenamefont {Chen},
  \citenamefont {Geng}, \citenamefont {Zhang} \emph {et~al.}}]{Wenyu_QPC}%
  \BibitemOpen
  \bibfield  {author} {\bibinfo {author} {\bibfnamefont {W.}~\bibnamefont
  {Song}}, \bibinfo {author} {\bibfnamefont {Y.}~\bibnamefont {Wang}}, \bibinfo
  {author} {\bibfnamefont {W.}~\bibnamefont {Miao}}, \bibinfo {author}
  {\bibfnamefont {Z.}~\bibnamefont {Yu}}, \bibinfo {author} {\bibfnamefont
  {Y.}~\bibnamefont {Gao}}, \bibinfo {author} {\bibfnamefont {R.}~\bibnamefont
  {Li}}, \bibinfo {author} {\bibfnamefont {S.}~\bibnamefont {Yang}}, \bibinfo
  {author} {\bibfnamefont {F.}~\bibnamefont {Chen}}, \bibinfo {author}
  {\bibfnamefont {Z.}~\bibnamefont {Geng}}, \bibinfo {author} {\bibfnamefont
  {Z.}~\bibnamefont {Zhang}}, \emph {et~al.},\ }\bibfield  {title} {\bibinfo
  {title} {Conductance quantization in {PbTe} nanowires},\ }\href
  {https://doi.org/10.1103/PhysRevB.108.045426} {\bibfield  {journal} {\bibinfo
   {journal} {Phys. Rev. B}\ }\textbf {\bibinfo {volume} {108}},\ \bibinfo
  {pages} {045426} (\bibinfo {year} {2023})}\BibitemShut {NoStop}%
\bibitem [{\citenamefont {Gao}\ \emph {et~al.}(2024{\natexlab{a}})\citenamefont
  {Gao}, \citenamefont {Song}, \citenamefont {Yang}, \citenamefont {Yu},
  \citenamefont {Li}, \citenamefont {Miao}, \citenamefont {Wang}, \citenamefont
  {Chen}, \citenamefont {Geng}, \citenamefont {Yang} \emph
  {et~al.}}]{Yichun_Gap}%
  \BibitemOpen
  \bibfield  {author} {\bibinfo {author} {\bibfnamefont {Y.}~\bibnamefont
  {Gao}}, \bibinfo {author} {\bibfnamefont {W.}~\bibnamefont {Song}}, \bibinfo
  {author} {\bibfnamefont {S.}~\bibnamefont {Yang}}, \bibinfo {author}
  {\bibfnamefont {Z.}~\bibnamefont {Yu}}, \bibinfo {author} {\bibfnamefont
  {R.}~\bibnamefont {Li}}, \bibinfo {author} {\bibfnamefont {W.}~\bibnamefont
  {Miao}}, \bibinfo {author} {\bibfnamefont {Y.}~\bibnamefont {Wang}}, \bibinfo
  {author} {\bibfnamefont {F.}~\bibnamefont {Chen}}, \bibinfo {author}
  {\bibfnamefont {Z.}~\bibnamefont {Geng}}, \bibinfo {author} {\bibfnamefont
  {L.}~\bibnamefont {Yang}}, \emph {et~al.},\ }\bibfield  {title} {\bibinfo
  {title} {Hard superconducting gap in {PbTe} nanowires},\ }\href
  {https://doi.org/10.1088/0256-307X/41/3/038502} {\bibfield  {journal}
  {\bibinfo  {journal} {Chinese Physics Letters}\ }\textbf {\bibinfo {volume}
  {41}},\ \bibinfo {pages} {038502} (\bibinfo {year}
  {2024}{\natexlab{a}})}\BibitemShut {NoStop}%
\bibitem [{\citenamefont {Wang}\ \emph {et~al.}(2023)\citenamefont {Wang},
  \citenamefont {Chen}, \citenamefont {Song}, \citenamefont {Geng},
  \citenamefont {Yu}, \citenamefont {Yang}, \citenamefont {Gao}, \citenamefont
  {Li}, \citenamefont {Yang}, \citenamefont {Miao} \emph {et~al.}}]{Yuhao_QPC}%
  \BibitemOpen
  \bibfield  {author} {\bibinfo {author} {\bibfnamefont {Y.}~\bibnamefont
  {Wang}}, \bibinfo {author} {\bibfnamefont {F.}~\bibnamefont {Chen}}, \bibinfo
  {author} {\bibfnamefont {W.}~\bibnamefont {Song}}, \bibinfo {author}
  {\bibfnamefont {Z.}~\bibnamefont {Geng}}, \bibinfo {author} {\bibfnamefont
  {Z.}~\bibnamefont {Yu}}, \bibinfo {author} {\bibfnamefont {L.}~\bibnamefont
  {Yang}}, \bibinfo {author} {\bibfnamefont {Y.}~\bibnamefont {Gao}}, \bibinfo
  {author} {\bibfnamefont {R.}~\bibnamefont {Li}}, \bibinfo {author}
  {\bibfnamefont {S.}~\bibnamefont {Yang}}, \bibinfo {author} {\bibfnamefont
  {W.}~\bibnamefont {Miao}}, \emph {et~al.},\ }\bibfield  {title} {\bibinfo
  {title} {Ballistic {PbTe} nanowire devices},\ }\href
  {https://doi.org/10.1021/acs.nanolett.3c03604} {\bibfield  {journal}
  {\bibinfo  {journal} {Nano Letters}\ }\textbf {\bibinfo {volume} {23}},\
  \bibinfo {pages} {11137} (\bibinfo {year} {2023})}\BibitemShut {NoStop}%
\bibitem [{\citenamefont {Li}\ \emph {et~al.}(2024)\citenamefont {Li},
  \citenamefont {Song}, \citenamefont {Miao}, \citenamefont {Yu}, \citenamefont
  {Wang}, \citenamefont {Yang}, \citenamefont {Gao}, \citenamefont {Wang},
  \citenamefont {Chen}, \citenamefont {Geng} \emph {et~al.}}]{Ruidong_PJJ}%
  \BibitemOpen
  \bibfield  {author} {\bibinfo {author} {\bibfnamefont {R.}~\bibnamefont
  {Li}}, \bibinfo {author} {\bibfnamefont {W.}~\bibnamefont {Song}}, \bibinfo
  {author} {\bibfnamefont {W.}~\bibnamefont {Miao}}, \bibinfo {author}
  {\bibfnamefont {Z.}~\bibnamefont {Yu}}, \bibinfo {author} {\bibfnamefont
  {Z.}~\bibnamefont {Wang}}, \bibinfo {author} {\bibfnamefont {S.}~\bibnamefont
  {Yang}}, \bibinfo {author} {\bibfnamefont {Y.}~\bibnamefont {Gao}}, \bibinfo
  {author} {\bibfnamefont {Y.}~\bibnamefont {Wang}}, \bibinfo {author}
  {\bibfnamefont {F.}~\bibnamefont {Chen}}, \bibinfo {author} {\bibfnamefont
  {Z.}~\bibnamefont {Geng}}, \emph {et~al.},\ }\bibfield  {title} {\bibinfo
  {title} {Selective-area-grown {PbTe-Pb} planar {J}osephson junctions for
  quantum devices},\ }\href {https://doi.org/10.1021/acs.nanolett.4c00900}
  {\bibfield  {journal} {\bibinfo  {journal} {Nano Letters}\ }\textbf {\bibinfo
  {volume} {24}},\ \bibinfo {pages} {4658} (\bibinfo {year}
  {2024})}\BibitemShut {NoStop}%
\bibitem [{\citenamefont {Gupta}\ \emph {et~al.}(2024)\citenamefont {Gupta},
  \citenamefont {Khade}, \citenamefont {Riggert}, \citenamefont {Shani},
  \citenamefont {Menning}, \citenamefont {Lueb}, \citenamefont {Jung},
  \citenamefont {M{\'e}lin}, \citenamefont {Bakkers},\ and\ \citenamefont
  {Pribiag}}]{Vlad_PbTe}%
  \BibitemOpen
  \bibfield  {author} {\bibinfo {author} {\bibfnamefont {M.}~\bibnamefont
  {Gupta}}, \bibinfo {author} {\bibfnamefont {V.}~\bibnamefont {Khade}},
  \bibinfo {author} {\bibfnamefont {C.}~\bibnamefont {Riggert}}, \bibinfo
  {author} {\bibfnamefont {L.}~\bibnamefont {Shani}}, \bibinfo {author}
  {\bibfnamefont {G.}~\bibnamefont {Menning}}, \bibinfo {author} {\bibfnamefont
  {P.~J.~H.}\ \bibnamefont {Lueb}}, \bibinfo {author} {\bibfnamefont
  {J.}~\bibnamefont {Jung}}, \bibinfo {author} {\bibfnamefont {R.}~\bibnamefont
  {M{\'e}lin}}, \bibinfo {author} {\bibfnamefont {E.~P. A.~M.}\ \bibnamefont
  {Bakkers}},\ and\ \bibinfo {author} {\bibfnamefont {V.~S.}\ \bibnamefont
  {Pribiag}},\ }\bibfield  {title} {\bibinfo {title} {Evidence for
  $\pi$-shifted cooper quartets and few-mode transport in pbte nanowire
  three-terminal josephson junctions},\ }\href
  {https://doi.org/10.1021/acs.nanolett.4c02414} {\bibfield  {journal}
  {\bibinfo  {journal} {Nano Letters}\ }\textbf {\bibinfo {volume} {24}},\
  \bibinfo {pages} {13903} (\bibinfo {year} {2024})}\BibitemShut {NoStop}%
\bibitem [{\citenamefont {Song}\ \emph {et~al.}(2025)\citenamefont {Song},
  \citenamefont {Yu}, \citenamefont {Wang}, \citenamefont {Gao}, \citenamefont
  {Li}, \citenamefont {Yang}, \citenamefont {Zhang}, \citenamefont {Geng},
  \citenamefont {Li}, \citenamefont {Wang} \emph {et~al.}}]{Wenyu_Disorder}%
  \BibitemOpen
  \bibfield  {author} {\bibinfo {author} {\bibfnamefont {W.}~\bibnamefont
  {Song}}, \bibinfo {author} {\bibfnamefont {Z.}~\bibnamefont {Yu}}, \bibinfo
  {author} {\bibfnamefont {Y.}~\bibnamefont {Wang}}, \bibinfo {author}
  {\bibfnamefont {Y.}~\bibnamefont {Gao}}, \bibinfo {author} {\bibfnamefont
  {Z.}~\bibnamefont {Li}}, \bibinfo {author} {\bibfnamefont {S.}~\bibnamefont
  {Yang}}, \bibinfo {author} {\bibfnamefont {S.}~\bibnamefont {Zhang}},
  \bibinfo {author} {\bibfnamefont {Z.}~\bibnamefont {Geng}}, \bibinfo {author}
  {\bibfnamefont {R.}~\bibnamefont {Li}}, \bibinfo {author} {\bibfnamefont
  {Z.}~\bibnamefont {Wang}}, \emph {et~al.},\ }\bibfield  {title} {\bibinfo
  {title} {Reducing disorder in {PbTe} nanowires for {M}ajorana research},\
  }\href {https://doi.org/10.1021/acs.nanolett.4c05708} {\bibfield  {journal}
  {\bibinfo  {journal} {Nano Letters}\ }\textbf {\bibinfo {volume} {25}},\
  \bibinfo {pages} {2350} (\bibinfo {year} {2025})}\BibitemShut {NoStop}%
\bibitem [{\citenamefont {Geng}\ \emph {et~al.}(2025)\citenamefont {Geng},
  \citenamefont {Chen}, \citenamefont {Gao}, \citenamefont {Yang},
  \citenamefont {Wang}, \citenamefont {Yang}, \citenamefont {Zhang},
  \citenamefont {Li}, \citenamefont {Song}, \citenamefont {Xu} \emph
  {et~al.}}]{PbTe_In}%
  \BibitemOpen
  \bibfield  {author} {\bibinfo {author} {\bibfnamefont {Z.}~\bibnamefont
  {Geng}}, \bibinfo {author} {\bibfnamefont {F.}~\bibnamefont {Chen}}, \bibinfo
  {author} {\bibfnamefont {Y.}~\bibnamefont {Gao}}, \bibinfo {author}
  {\bibfnamefont {L.}~\bibnamefont {Yang}}, \bibinfo {author} {\bibfnamefont
  {Y.}~\bibnamefont {Wang}}, \bibinfo {author} {\bibfnamefont {S.}~\bibnamefont
  {Yang}}, \bibinfo {author} {\bibfnamefont {S.}~\bibnamefont {Zhang}},
  \bibinfo {author} {\bibfnamefont {Z.}~\bibnamefont {Li}}, \bibinfo {author}
  {\bibfnamefont {W.}~\bibnamefont {Song}}, \bibinfo {author} {\bibfnamefont
  {J.}~\bibnamefont {Xu}}, \emph {et~al.},\ }\bibfield  {title} {\bibinfo
  {title} {Enhanced superconductivity in {PbTe-In} hybrids},\ }\href
  {https://doi.org/10.1103/8czc-tn4z} {\bibfield  {journal} {\bibinfo
  {journal} {Phys. Rev. Mater.}\ }\textbf {\bibinfo {volume} {9}},\ \bibinfo
  {pages} {084802} (\bibinfo {year} {2025})}\BibitemShut {NoStop}%
\bibitem [{\citenamefont {Wang}\ \emph {et~al.}(2024)\citenamefont {Wang},
  \citenamefont {Song}, \citenamefont {Cao}, \citenamefont {Yu}, \citenamefont
  {Yang}, \citenamefont {Li}, \citenamefont {Gao}, \citenamefont {Li},
  \citenamefont {Chen}, \citenamefont {Geng} \emph
  {et~al.}}]{Yuhao_degeneracy}%
  \BibitemOpen
  \bibfield  {author} {\bibinfo {author} {\bibfnamefont {Y.}~\bibnamefont
  {Wang}}, \bibinfo {author} {\bibfnamefont {W.}~\bibnamefont {Song}}, \bibinfo
  {author} {\bibfnamefont {Z.}~\bibnamefont {Cao}}, \bibinfo {author}
  {\bibfnamefont {Z.}~\bibnamefont {Yu}}, \bibinfo {author} {\bibfnamefont
  {S.}~\bibnamefont {Yang}}, \bibinfo {author} {\bibfnamefont {Z.}~\bibnamefont
  {Li}}, \bibinfo {author} {\bibfnamefont {Y.}~\bibnamefont {Gao}}, \bibinfo
  {author} {\bibfnamefont {R.}~\bibnamefont {Li}}, \bibinfo {author}
  {\bibfnamefont {F.}~\bibnamefont {Chen}}, \bibinfo {author} {\bibfnamefont
  {Z.}~\bibnamefont {Geng}}, \emph {et~al.},\ }\bibfield  {title} {\bibinfo
  {title} {Gate-tunable subband degeneracy in semiconductor nanowires},\ }\href
  {https://doi.org/10.1073/pnas.2406884121} {\bibfield  {journal} {\bibinfo
  {journal} {Proceedings of the National Academy of Sciences}\ }\textbf
  {\bibinfo {volume} {121}},\ \bibinfo {pages} {e2406884121} (\bibinfo {year}
  {2024})}\BibitemShut {NoStop}%
\bibitem [{\citenamefont {Gao}\ \emph {et~al.}(2024{\natexlab{b}})\citenamefont
  {Gao}, \citenamefont {Song}, \citenamefont {Yu}, \citenamefont {Yang},
  \citenamefont {Wang}, \citenamefont {Li}, \citenamefont {Chen}, \citenamefont
  {Geng}, \citenamefont {Yang}, \citenamefont {Xu} \emph
  {et~al.}}]{Yichun_SQUID}%
  \BibitemOpen
  \bibfield  {author} {\bibinfo {author} {\bibfnamefont {Y.}~\bibnamefont
  {Gao}}, \bibinfo {author} {\bibfnamefont {W.}~\bibnamefont {Song}}, \bibinfo
  {author} {\bibfnamefont {Z.}~\bibnamefont {Yu}}, \bibinfo {author}
  {\bibfnamefont {S.}~\bibnamefont {Yang}}, \bibinfo {author} {\bibfnamefont
  {Y.}~\bibnamefont {Wang}}, \bibinfo {author} {\bibfnamefont {R.}~\bibnamefont
  {Li}}, \bibinfo {author} {\bibfnamefont {F.}~\bibnamefont {Chen}}, \bibinfo
  {author} {\bibfnamefont {Z.}~\bibnamefont {Geng}}, \bibinfo {author}
  {\bibfnamefont {L.}~\bibnamefont {Yang}}, \bibinfo {author} {\bibfnamefont
  {J.}~\bibnamefont {Xu}}, \emph {et~al.},\ }\bibfield  {title} {\bibinfo
  {title} {{SQUID} oscillations in {PbTe} nanowire networks},\ }\href
  {https://doi.org/10.1103/PhysRevB.110.045405} {\bibfield  {journal} {\bibinfo
   {journal} {Phys. Rev. B}\ }\textbf {\bibinfo {volume} {110}},\ \bibinfo
  {pages} {045405} (\bibinfo {year} {2024}{\natexlab{b}})}\BibitemShut
  {NoStop}%
\bibitem [{\citenamefont {Gao}\ \emph {et~al.}(2025)\citenamefont {Gao},
  \citenamefont {Song}, \citenamefont {Wang}, \citenamefont {Geng},
  \citenamefont {Cao}, \citenamefont {Yu}, \citenamefont {Yang}, \citenamefont
  {Xu}, \citenamefont {Chen}, \citenamefont {Li} \emph
  {et~al.}}]{Quantized_Andreev}%
  \BibitemOpen
  \bibfield  {author} {\bibinfo {author} {\bibfnamefont {Y.}~\bibnamefont
  {Gao}}, \bibinfo {author} {\bibfnamefont {W.}~\bibnamefont {Song}}, \bibinfo
  {author} {\bibfnamefont {Y.}~\bibnamefont {Wang}}, \bibinfo {author}
  {\bibfnamefont {Z.}~\bibnamefont {Geng}}, \bibinfo {author} {\bibfnamefont
  {Z.}~\bibnamefont {Cao}}, \bibinfo {author} {\bibfnamefont {Z.}~\bibnamefont
  {Yu}}, \bibinfo {author} {\bibfnamefont {S.}~\bibnamefont {Yang}}, \bibinfo
  {author} {\bibfnamefont {J.}~\bibnamefont {Xu}}, \bibinfo {author}
  {\bibfnamefont {F.}~\bibnamefont {Chen}}, \bibinfo {author} {\bibfnamefont
  {Z.}~\bibnamefont {Li}}, \emph {et~al.},\ }\bibfield  {title} {\bibinfo
  {title} {Quantized {A}ndreev conductance in semiconductor nanowires},\ }\href
  {https://doi.org/10.1103/bwp9-7dsd} {\bibfield  {journal} {\bibinfo
  {journal} {Phys. Rev. Appl.}\ }\textbf {\bibinfo {volume} {23}},\ \bibinfo
  {pages} {L061004} (\bibinfo {year} {2025})}\BibitemShut {NoStop}%
\bibitem [{\citenamefont {Li}\ \emph {et~al.}(2025)\citenamefont {Li},
  \citenamefont {Song}, \citenamefont {Zhang}, \citenamefont {Wang},
  \citenamefont {Wang}, \citenamefont {Yu}, \citenamefont {Li}, \citenamefont
  {Yan}, \citenamefont {Xu}, \citenamefont {Gao} \emph
  {et~al.}}]{Zonglin_Anisotropy}%
  \BibitemOpen
  \bibfield  {author} {\bibinfo {author} {\bibfnamefont {Z.}~\bibnamefont
  {Li}}, \bibinfo {author} {\bibfnamefont {W.}~\bibnamefont {Song}}, \bibinfo
  {author} {\bibfnamefont {S.}~\bibnamefont {Zhang}}, \bibinfo {author}
  {\bibfnamefont {Y.}~\bibnamefont {Wang}}, \bibinfo {author} {\bibfnamefont
  {Z.}~\bibnamefont {Wang}}, \bibinfo {author} {\bibfnamefont {Z.}~\bibnamefont
  {Yu}}, \bibinfo {author} {\bibfnamefont {R.}~\bibnamefont {Li}}, \bibinfo
  {author} {\bibfnamefont {Z.}~\bibnamefont {Yan}}, \bibinfo {author}
  {\bibfnamefont {J.}~\bibnamefont {Xu}}, \bibinfo {author} {\bibfnamefont
  {Y.}~\bibnamefont {Gao}}, \emph {et~al.},\ }\bibfield  {title} {\bibinfo
  {title} {Anisotropy of {PbTe} nanowires with and without a superconductor},\
  }\href {https://doi.org/10.1103/PhysRevB.111.195416} {\bibfield  {journal}
  {\bibinfo  {journal} {Phys. Rev. B}\ }\textbf {\bibinfo {volume} {111}},\
  \bibinfo {pages} {195416} (\bibinfo {year} {2025})}\BibitemShut {NoStop}%
\bibitem [{\citenamefont {Winkler}\ \emph {et~al.}(2017)\citenamefont
  {Winkler}, \citenamefont {Varjas}, \citenamefont {Skolasinski}, \citenamefont
  {Soluyanov}, \citenamefont {Troyer},\ and\ \citenamefont
  {Wimmer}}]{Wimmer2017Orbital}%
  \BibitemOpen
  \bibfield  {author} {\bibinfo {author} {\bibfnamefont {G.~W.}\ \bibnamefont
  {Winkler}}, \bibinfo {author} {\bibfnamefont {D.}~\bibnamefont {Varjas}},
  \bibinfo {author} {\bibfnamefont {R.}~\bibnamefont {Skolasinski}}, \bibinfo
  {author} {\bibfnamefont {A.~A.}\ \bibnamefont {Soluyanov}}, \bibinfo {author}
  {\bibfnamefont {M.}~\bibnamefont {Troyer}},\ and\ \bibinfo {author}
  {\bibfnamefont {M.}~\bibnamefont {Wimmer}},\ }\bibfield  {title} {\bibinfo
  {title} {Orbital contributions to the electron g factor in semiconductor
  nanowires},\ }\href {https://doi.org/10.1103/PhysRevLett.119.037701}
  {\bibfield  {journal} {\bibinfo  {journal} {Physical Review Letters}\
  }\textbf {\bibinfo {volume} {119}},\ \bibinfo {pages} {037701} (\bibinfo
  {year} {2017})}\BibitemShut {NoStop}%
\bibitem [{\citenamefont {Vaitiekėnas}\ \emph {et~al.}(2020)\citenamefont
  {Vaitiekėnas}, \citenamefont {Winkler}, \citenamefont {van Heck},
  \citenamefont {Karzig}, \citenamefont {Deng}, \citenamefont {Flensberg},
  \citenamefont {Glazman}, \citenamefont {Nayak}, \citenamefont {Krogstrup},
  \citenamefont {Lutchyn},\ and\ \citenamefont {Marcus}}]{full_shell}%
  \BibitemOpen
  \bibfield  {author} {\bibinfo {author} {\bibfnamefont {S.}~\bibnamefont
  {Vaitiekėnas}}, \bibinfo {author} {\bibfnamefont {G.~W.}\ \bibnamefont
  {Winkler}}, \bibinfo {author} {\bibfnamefont {B.}~\bibnamefont {van Heck}},
  \bibinfo {author} {\bibfnamefont {T.}~\bibnamefont {Karzig}}, \bibinfo
  {author} {\bibfnamefont {M.-T.}\ \bibnamefont {Deng}}, \bibinfo {author}
  {\bibfnamefont {K.}~\bibnamefont {Flensberg}}, \bibinfo {author}
  {\bibfnamefont {L.~I.}\ \bibnamefont {Glazman}}, \bibinfo {author}
  {\bibfnamefont {C.}~\bibnamefont {Nayak}}, \bibinfo {author} {\bibfnamefont
  {P.}~\bibnamefont {Krogstrup}}, \bibinfo {author} {\bibfnamefont {R.~M.}\
  \bibnamefont {Lutchyn}},\ and\ \bibinfo {author} {\bibfnamefont {C.~M.}\
  \bibnamefont {Marcus}},\ }\bibfield  {title} {\bibinfo {title} {Flux-induced
  topological superconductivity in full-shell nanowires},\ }\href
  {https://doi.org/10.1126/science.aav3392} {\bibfield  {journal} {\bibinfo
  {journal} {Science}\ }\textbf {\bibinfo {volume} {367}},\ \bibinfo {pages}
  {eaav3392} (\bibinfo {year} {2020})}\BibitemShut {NoStop}%
\bibitem [{\citenamefont {Liu}\ \emph {et~al.}(2012)\citenamefont {Liu},
  \citenamefont {Potter}, \citenamefont {Law},\ and\ \citenamefont
  {Lee}}]{Patrick_Lee_disorder_2012}%
  \BibitemOpen
  \bibfield  {author} {\bibinfo {author} {\bibfnamefont {J.}~\bibnamefont
  {Liu}}, \bibinfo {author} {\bibfnamefont {A.~C.}\ \bibnamefont {Potter}},
  \bibinfo {author} {\bibfnamefont {K.~T.}\ \bibnamefont {Law}},\ and\ \bibinfo
  {author} {\bibfnamefont {P.~A.}\ \bibnamefont {Lee}},\ }\bibfield  {title}
  {\bibinfo {title} {Zero-bias peaks in the tunneling conductance of
  spin-orbit-coupled superconducting wires with and without {M}ajorana
  end-states},\ }\href {https://doi.org/10.1103/PhysRevLett.109.267002}
  {\bibfield  {journal} {\bibinfo  {journal} {Phys. Rev. Lett.}\ }\textbf
  {\bibinfo {volume} {109}},\ \bibinfo {pages} {267002} (\bibinfo {year}
  {2012})}\BibitemShut {NoStop}%
\bibitem [{\citenamefont {Prada}\ \emph {et~al.}(2012)\citenamefont {Prada},
  \citenamefont {San-Jose},\ and\ \citenamefont {Aguado}}]{Prada2012}%
  \BibitemOpen
  \bibfield  {author} {\bibinfo {author} {\bibfnamefont {E.}~\bibnamefont
  {Prada}}, \bibinfo {author} {\bibfnamefont {P.}~\bibnamefont {San-Jose}},\
  and\ \bibinfo {author} {\bibfnamefont {R.}~\bibnamefont {Aguado}},\
  }\bibfield  {title} {\bibinfo {title} {Transport spectroscopy of {NS}
  nanowire junctions with {M}ajorana fermions},\ }\href
  {https://doi.org/10.1103/PhysRevB.86.180503} {\bibfield  {journal} {\bibinfo
  {journal} {Physical Review B}\ }\textbf {\bibinfo {volume} {86}},\ \bibinfo
  {pages} {180503} (\bibinfo {year} {2012})}\BibitemShut {NoStop}%
\bibitem [{\citenamefont {Rainis}\ \emph {et~al.}(2013)\citenamefont {Rainis},
  \citenamefont {Trifunovic}, \citenamefont {Klinovaja},\ and\ \citenamefont
  {Loss}}]{Loss2013ZBP}%
  \BibitemOpen
  \bibfield  {author} {\bibinfo {author} {\bibfnamefont {D.}~\bibnamefont
  {Rainis}}, \bibinfo {author} {\bibfnamefont {L.}~\bibnamefont {Trifunovic}},
  \bibinfo {author} {\bibfnamefont {J.}~\bibnamefont {Klinovaja}},\ and\
  \bibinfo {author} {\bibfnamefont {D.}~\bibnamefont {Loss}},\ }\bibfield
  {title} {\bibinfo {title} {Towards a realistic transport modeling in a
  superconducting nanowire with {M}ajorana fermions},\ }\href
  {https://doi.org/10.1103/PhysRevB.87.024515} {\bibfield  {journal} {\bibinfo
  {journal} {Physical Review B}\ }\textbf {\bibinfo {volume} {87}},\ \bibinfo
  {pages} {024515} (\bibinfo {year} {2013})}\BibitemShut {NoStop}%
\bibitem [{\citenamefont {Lee}\ \emph {et~al.}(2014)\citenamefont {Lee},
  \citenamefont {Jiang}, \citenamefont {Houzet}, \citenamefont {Aguado},
  \citenamefont {Lieber},\ and\ \citenamefont {De~Franceschi}}]{Silvano2014}%
  \BibitemOpen
  \bibfield  {author} {\bibinfo {author} {\bibfnamefont {E.~J.}\ \bibnamefont
  {Lee}}, \bibinfo {author} {\bibfnamefont {X.}~\bibnamefont {Jiang}}, \bibinfo
  {author} {\bibfnamefont {M.}~\bibnamefont {Houzet}}, \bibinfo {author}
  {\bibfnamefont {R.}~\bibnamefont {Aguado}}, \bibinfo {author} {\bibfnamefont
  {C.~M.}\ \bibnamefont {Lieber}},\ and\ \bibinfo {author} {\bibfnamefont
  {S.}~\bibnamefont {De~Franceschi}},\ }\bibfield  {title} {\bibinfo {title}
  {Spin-resolved {A}ndreev levels and parity crossings in hybrid
  superconductor--semiconductor nanostructures},\ }\href
  {https://doi.org/10.1038/nnano.2013.267} {\bibfield  {journal} {\bibinfo
  {journal} {Nature Nanotechnology}\ }\textbf {\bibinfo {volume} {9}},\
  \bibinfo {pages} {79} (\bibinfo {year} {2014})}\BibitemShut {NoStop}%
\bibitem [{\citenamefont {Pan}\ and\ \citenamefont
  {Das~Sarma}(2020)}]{GoodBadUgly}%
  \BibitemOpen
  \bibfield  {author} {\bibinfo {author} {\bibfnamefont {H.}~\bibnamefont
  {Pan}}\ and\ \bibinfo {author} {\bibfnamefont {S.}~\bibnamefont
  {Das~Sarma}},\ }\bibfield  {title} {\bibinfo {title} {Physical mechanisms for
  zero-bias conductance peaks in {M}ajorana nanowires},\ }\href
  {https://doi.org/10.1103/PhysRevResearch.2.013377} {\bibfield  {journal}
  {\bibinfo  {journal} {Phys. Rev. Research}\ }\textbf {\bibinfo {volume}
  {2}},\ \bibinfo {pages} {013377} (\bibinfo {year} {2020})}\BibitemShut
  {NoStop}%
\bibitem [{\citenamefont {Ahn}\ \emph {et~al.}(2021)\citenamefont {Ahn},
  \citenamefont {Pan}, \citenamefont {Woods}, \citenamefont {Stanescu},\ and\
  \citenamefont {Das~Sarma}}]{DasSarma_estimate}%
  \BibitemOpen
  \bibfield  {author} {\bibinfo {author} {\bibfnamefont {S.}~\bibnamefont
  {Ahn}}, \bibinfo {author} {\bibfnamefont {H.}~\bibnamefont {Pan}}, \bibinfo
  {author} {\bibfnamefont {B.}~\bibnamefont {Woods}}, \bibinfo {author}
  {\bibfnamefont {T.~D.}\ \bibnamefont {Stanescu}},\ and\ \bibinfo {author}
  {\bibfnamefont {S.}~\bibnamefont {Das~Sarma}},\ }\bibfield  {title} {\bibinfo
  {title} {Estimating disorder and its adverse effects in semiconductor
  majorana nanowires},\ }\href
  {https://doi.org/10.1103/PhysRevMaterials.5.124602} {\bibfield  {journal}
  {\bibinfo  {journal} {Phys. Rev. Materials}\ }\textbf {\bibinfo {volume}
  {5}},\ \bibinfo {pages} {124602} (\bibinfo {year} {2021})}\BibitemShut
  {NoStop}%
\bibitem [{\citenamefont {Zeng}\ \emph {et~al.}(2022)\citenamefont {Zeng},
  \citenamefont {Sharma}, \citenamefont {Tewari},\ and\ \citenamefont
  {Stanescu}}]{Tudor2021Disorder}%
  \BibitemOpen
  \bibfield  {author} {\bibinfo {author} {\bibfnamefont {C.}~\bibnamefont
  {Zeng}}, \bibinfo {author} {\bibfnamefont {G.}~\bibnamefont {Sharma}},
  \bibinfo {author} {\bibfnamefont {S.}~\bibnamefont {Tewari}},\ and\ \bibinfo
  {author} {\bibfnamefont {T.}~\bibnamefont {Stanescu}},\ }\bibfield  {title}
  {\bibinfo {title} {Partially separated {M}ajorana modes in a disordered
  medium},\ }\href {https://doi.org/10.1103/PhysRevB.105.205122} {\bibfield
  {journal} {\bibinfo  {journal} {Phys. Rev. B}\ }\textbf {\bibinfo {volume}
  {105}},\ \bibinfo {pages} {205122} (\bibinfo {year} {2022})}\BibitemShut
  {NoStop}%
\bibitem [{\citenamefont {Hess}\ \emph {et~al.}(2023)\citenamefont {Hess},
  \citenamefont {Legg}, \citenamefont {Loss},\ and\ \citenamefont
  {Klinovaja}}]{Loss_Andreev_band}%
  \BibitemOpen
  \bibfield  {author} {\bibinfo {author} {\bibfnamefont {R.}~\bibnamefont
  {Hess}}, \bibinfo {author} {\bibfnamefont {H.~F.}\ \bibnamefont {Legg}},
  \bibinfo {author} {\bibfnamefont {D.}~\bibnamefont {Loss}},\ and\ \bibinfo
  {author} {\bibfnamefont {J.}~\bibnamefont {Klinovaja}},\ }\bibfield  {title}
  {\bibinfo {title} {Trivial {A}ndreev band mimicking topological bulk gap
  reopening in the nonlocal conductance of long {R}ashba nanowires},\ }\href
  {https://doi.org/10.1103/PhysRevLett.130.207001} {\bibfield  {journal}
  {\bibinfo  {journal} {Phys. Rev. Lett.}\ }\textbf {\bibinfo {volume} {130}},\
  \bibinfo {pages} {207001} (\bibinfo {year} {2023})}\BibitemShut {NoStop}%
\end{thebibliography}%

\newpage

\onecolumngrid

\newpage
\includepdf[pages=1]{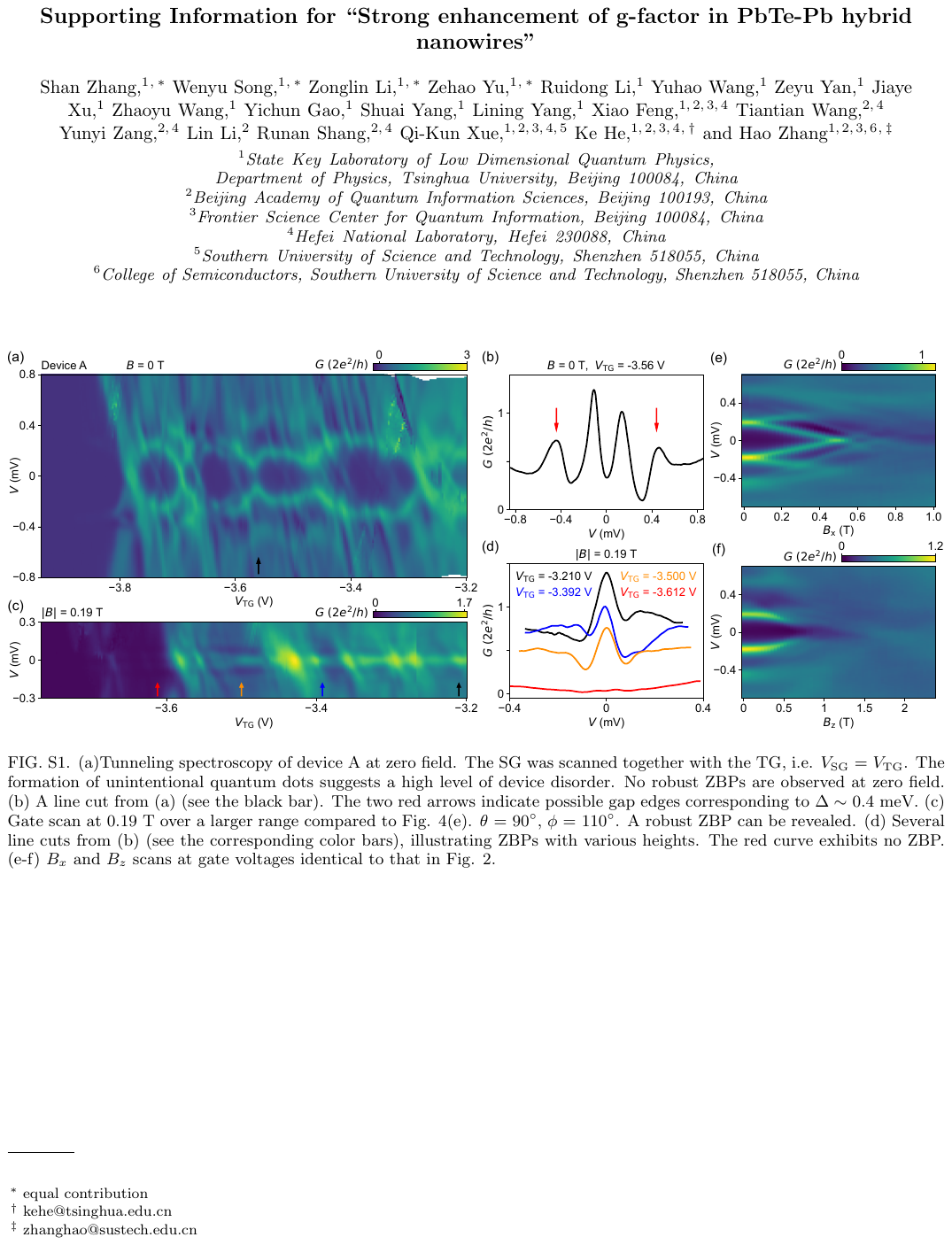}
\includepdf[pages=2]{PbTe_g_SM.pdf}
\includepdf[pages=3]{PbTe_g_SM.pdf}
\includepdf[pages=4]{PbTe_g_SM.pdf}
\end{document}